\documentclass{article}

\usepackage{arxiv}

\usepackage[utf8]{inputenc} 
\usepackage[T1]{fontenc}    
\usepackage{hyperref}       
\hypersetup{
    colorlinks=true,
    linkcolor=black,
    urlcolor=blue,
    citecolor=blue
}

\usepackage{url}            
\usepackage{booktabs}       
\usepackage{amsfonts}       
\usepackage{nicefrac}       
\usepackage{microtype}      
\usepackage{lipsum}		
\usepackage{graphicx}
\usepackage{natbib}
\usepackage{pdflscape}
\usepackage{rotating}
\usepackage[table]{xcolor}
\usepackage{colortbl}
\usepackage{caption}  
\usepackage{threeparttable}
\usepackage{wrapfig}
\usepackage{doi}
\usepackage{tocloft}

\title{Technical Overview of Recent Developments in Small Modular Reactors in the United States}

\author{{\hspace{1mm}Yifan Sun} \\
	Kyoto University\\
	\And
    {\hspace{1mm}Ken Kurosaki} \\
	Kyoto University\\
}

\renewcommand{\headeright}{Technical Review}
\renewcommand{\undertitle}{Technical Review}
\renewcommand{\shorttitle}{}

\begin{document}
\maketitle

{
\tableofcontents
}

\newpage

\section{Introduction}
\subsection{Overview of SMRs}
\subsubsection*{What are SMRs?}
Small Modular Reactors (SMRs) are a family of advanced nuclear fission reactors that differ from conventional reactors in their smaller size and modular design. Each SMR unit typically has a designed power capacity of up to 300 MWe [1], approximately one-third that of traditional reactors (e.g., the Shimane-3 ABWR, which is under construction, has a designed net capacity of 1,325 MWe [2]). The exact definition of an SMR varies across organizations; some classify microreactor (below 20 MWe) as a separate category. In this report, we follow the definitions set by the International Atomic Energy Agency (IAEA) and Nuclear Energy Agency (NEA), which consider microreactor a subcategory of SMRs.

One of the main advantages of SMRs is their smaller footprint, making them ideal for locations where full-scale power plants are impractical. According to the data compiled by Hannah Ritchie and published on Our World in Data, nuclear power is already the most land-efficient energy source, requiring just 0.3 m\textsuperscript{2} of land per MWh generated annually [3]. SMRs, such as NuScale’s 12-module VOYGR SMR plant, are expected to achieve even greater energy density, generating 924 MWe on approximately 0.16 km\textsuperscript{2} of land [4]. Similarly, the BWRX-300 SMR by GE-Hitachi is advertised as capable of powering 300,000 homes with a plant the size of a football field [5]. Microreactors, being even more compact and transportable by truck, are envisioned for remote power generation and as diesel replacements for emergency backup power and disaster relief.

The modular nature of SMRs allows their components to be manufactured off-site and later assembled on location, making deployment more flexible and less site-dependent. Their scalable power output ensures they do not overload existing grids, while multi-module designs can expand capacity as needed [6]. Additionally, SMRs can promote nuclear integration into industrial processes by cogenerating both electricity and high-temperature heat. For example, Molten Salt Reactors and High-Temperature Gas-Cooled Reactors can provide the high temperatures needed for steelmaking and hydrogen production via high-temperature electrolysis, while reducing overall CO\textsubscript{2} emissions [7].

Due to their smaller core size, SMRs have lower power output, a reduced nuclear fuel inventory, which lessens the need for extensive on-site radiation shielding and reduces accident consequences. They also incorporate passive safety features, such as gravity- and convection-driven cooling systems, to eliminate reliance on external power or active human intervention for shutdown and cooling. Further safety improvements, such as design simplifications in reducing vessel penetrations, reduce potential pipe breach risks [8]. Additionally, some SMR designs aim for extended refueling cycles beyond the 18–24 months typical of conventional nuclear plants [9]. For instance, Canada’s sodium-cooled ARC-100 fast reactor is designed for a 20-year refueling period, while designs like X-energy's Xe-100, a high-temperature gas-cooled reactor, operate with continuous refueling [10].

\subsubsection*{Design Classifications}
Generally speaking, SMR designs can be classified as either being variations of existing Gen II and Gen III/III+ LWR-based reactors or more advanced GEN-IV reactors using coolants such as molten salt and liquid metals [8]. LWR-based SMRs are more mature in terms of both technological and licensing readiness as they can draw from decades of operating experiences. 

Based on key design characteristics such as fuel type, coolant, and intended applications, existing SMR designs can be grouped into the following categories:
\begin{itemize}
    \item Land-based water-cooled SMRs
    \item Marine-based water-cooled SMRs
    \item High-temperature gas-cooled SMRs
    \item Liquid metal-cooled fast neutron SMRs
    \item Molten salt SMRs
    \item Microreactors
\end{itemize}

\begin{figure}[h]
    \centering
    \includegraphics[width=0.9\linewidth]{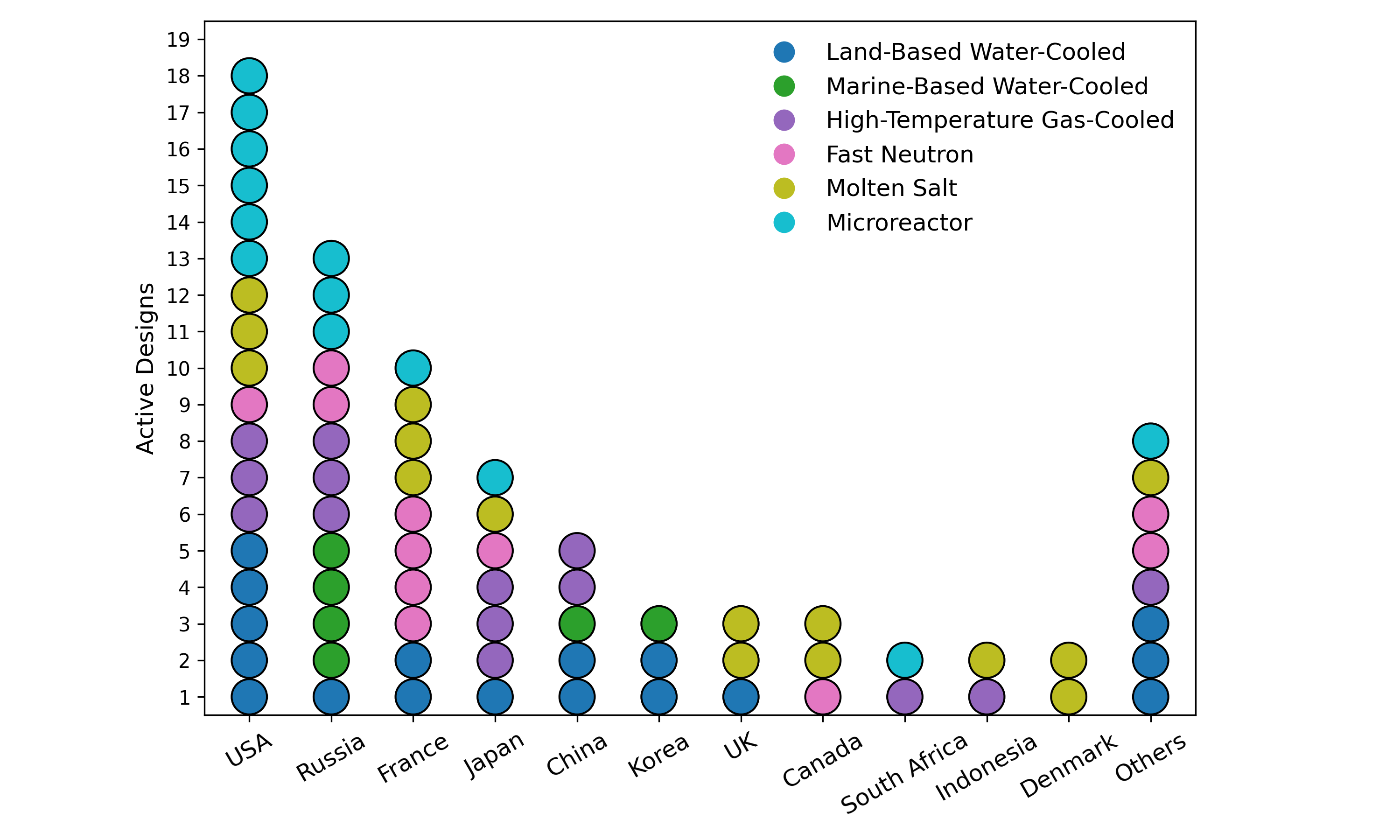}
    \caption{Active SMR designs and their classifications by country. Countries with only one active design, such as Argentina, Saudi Arabia, and Switzerland, are grouped under "Others."}
    \label{fig:world-wide-smr}
\end{figure}

The number of active SMR designs recorded by different organizations varies slightly due to differing selection criteria. The latest 2024 edition of IAEA's Small Modular Reactor Technology Catalogue\footnote{The contents of the catalogue have not been officially reviewed by the IAEA, and the catalogue is therefore not considered an official IAEA publication} cites 70 active SMR designs that demonstrate continuous development [10], whereas the 2024 NEA Small Modular Reactor Dashboard report includes only 56 out of 98 identified SMR technologies, selecting those with sufficient publicly available information [6]. Figure \ref{fig:world-wide-smr} illustrates the number of active and sustainable SMR designs for major countries, as listed in the 2024 IAEA Small Modular Reactor Technology Catalogue [10].

\begin{figure}[h]
    \centering
    \includegraphics[width=0.9\linewidth]{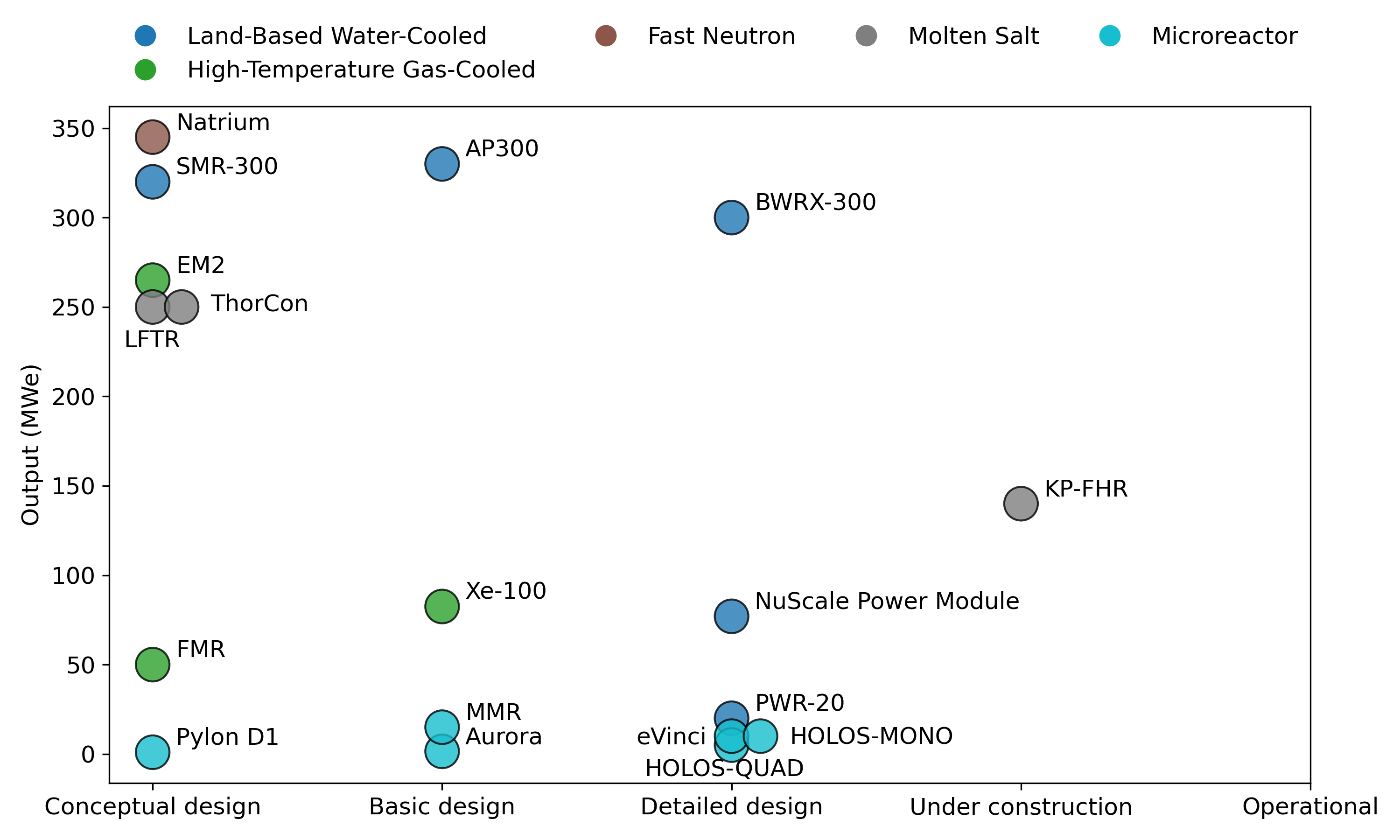}
    \caption{Development status of 18 active SMR designs in the United States, based on the 2024 edition of the Small Modular Reactor Technology Catalogue. eVinci is classified as "under development" in the catalogue but labeled here as "detailed design" for category consistency.}
    \label{fig:us-smr-designs}
\end{figure}

As of January 2025, the United States leads with the largest number of 18 active SMR designs, as listed in Figures \ref{fig:world-wide-smr} and \ref{fig:us-smr-designs}. This is followed by Russia, which has a strong focus on marine-based water-cooled SMRs. Among the designs shown in Figure \ref{fig:world-wide-smr}, only three high-temperature gas-cooled reactors, HTR-10 (China), HTR-PM (China), and HTTR (Japan), are operational. Additionally, two units of the marine-based water-cooled KLT-40S reactor are deployed on the Russian Akademik Lomonosov floating nuclear power plant, each with a 35 MWe output capacity [7].

Several other designs are currently listed as under construction by the IAEA [7], including:
\begin{itemize}
    \item CAREM-25\footnote{Construction was reported to be halted due to layoffs in Sept. 2024, according to the Buenos Aires Herald.} (PWR, Argentina)
    \item ACP100 (PWR, China)
    \item BREST-OD-300 (Fast neutron, Russia)
    \item KP-FHR (Molten salt, USA)
\end{itemize}

As of January 2025, the IAEA lists 18 U.S. SMRs as active and sustainable designs, as shown in Figure \ref{fig:us-smr-designs}. 

\subsection{Historical Context and Current Status}
Focusing on the United States, the initial concept of small nuclear reactors can be traced back to the Army Nuclear Power Program (1954–1977), during which the U.S. Army developed eight transportable reactors for remote operations [11,12]. While some considered these designs unreliable and expensive, they contributed to the advancement of the U.S. nuclear program for years to come [13].

By 1997, renewed interest in new reactor designs emerged, driven by the need for higher efficiency, lower costs, and improved safety, particularly in locations where large reactors were not feasible. Between 2000 and 2003, the Multi-Application Small Light Water Reactor project funded by the Department of Energy (DOE) led to the design of a 45 MW reactor, which later became the foundation of NuScale Power, founded in 2007 [14].

Around the same time, in 2009, Babcock \& Wilcox announced its 125 MW mPower reactor, which later won the DOE's Small Modular Reactor Licensing Technical Support Program in 2012, beating Westinghouse, NuScale Power, and Holtec. However, the mPower project was ultimately terminated in 2017 due to insufficient funding [15,16,17]. In that same year, NuScale Power became the first U.S. company to submit a Design Certification application to the Nuclear Regulatory Commission (NRC) for its NuScale US600 Design, which was later approved in 2020 and certified in 2023 [18,19,20]. Although NuScale Power initially planned to construct SMRs at Idaho National Laboratory with the Utah Associated Municipal Power Systems, the project was canceled in November 2023 [21]. 

\begin{figure}[h]
    \centering
    \includegraphics[width=\linewidth]{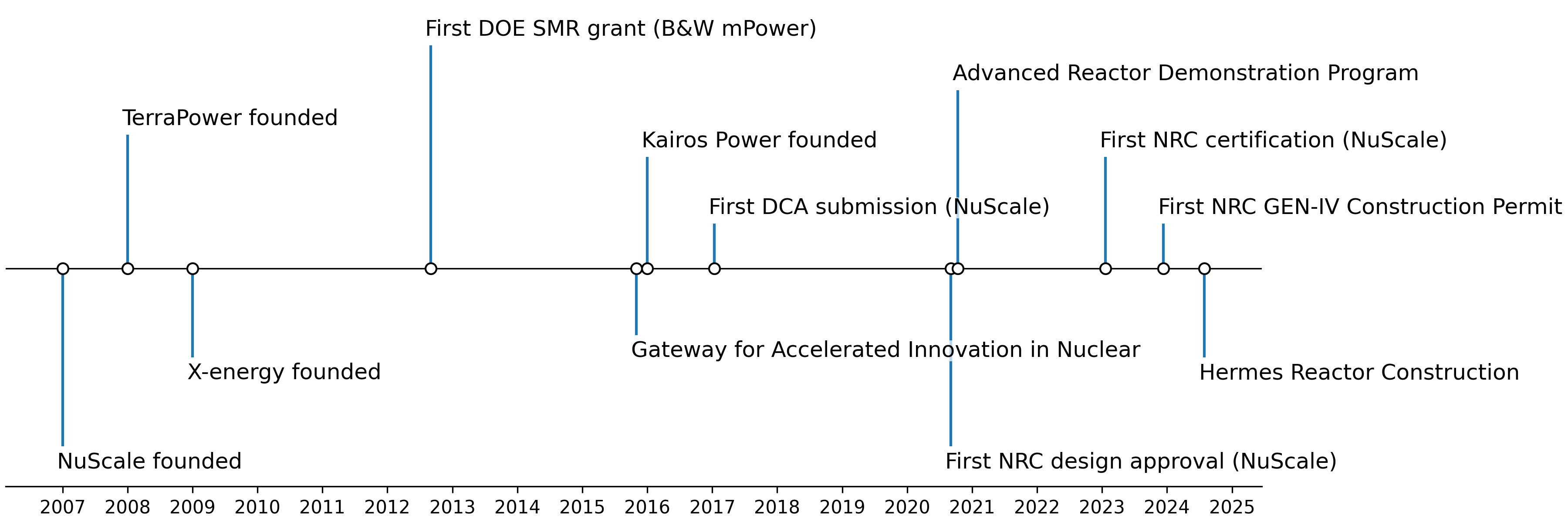}
    \caption{A simplified timeline highlighting some of the U.S. SMR development milestones in the last 20 years.}
    \label{fig:smr-timeline}
\end{figure}

Despite some earlier setbacks, recent years have seen progress in the development and certification of SMRs. For example, in March 2024, TerraPower submitted its Construction Permit application, and construction of the non-nuclear portion of the Natrium project began in Wyoming in June [22,23]. In July, Kairos Power began construction of its low-power Hermes Reactor, a demonstration project for the development of its molten salt KP-FHR reactor [24]. A simplified timeline highlighting selected key U.S. SMR development milestones is shown in Figure \ref{fig:smr-timeline}.

\subsection{Roles in the U.S. Energy Mix}

\begin{wrapfigure}{r}{0.5\textwidth}
    \vspace{-2mm}
    \includegraphics[width=0.95\linewidth]{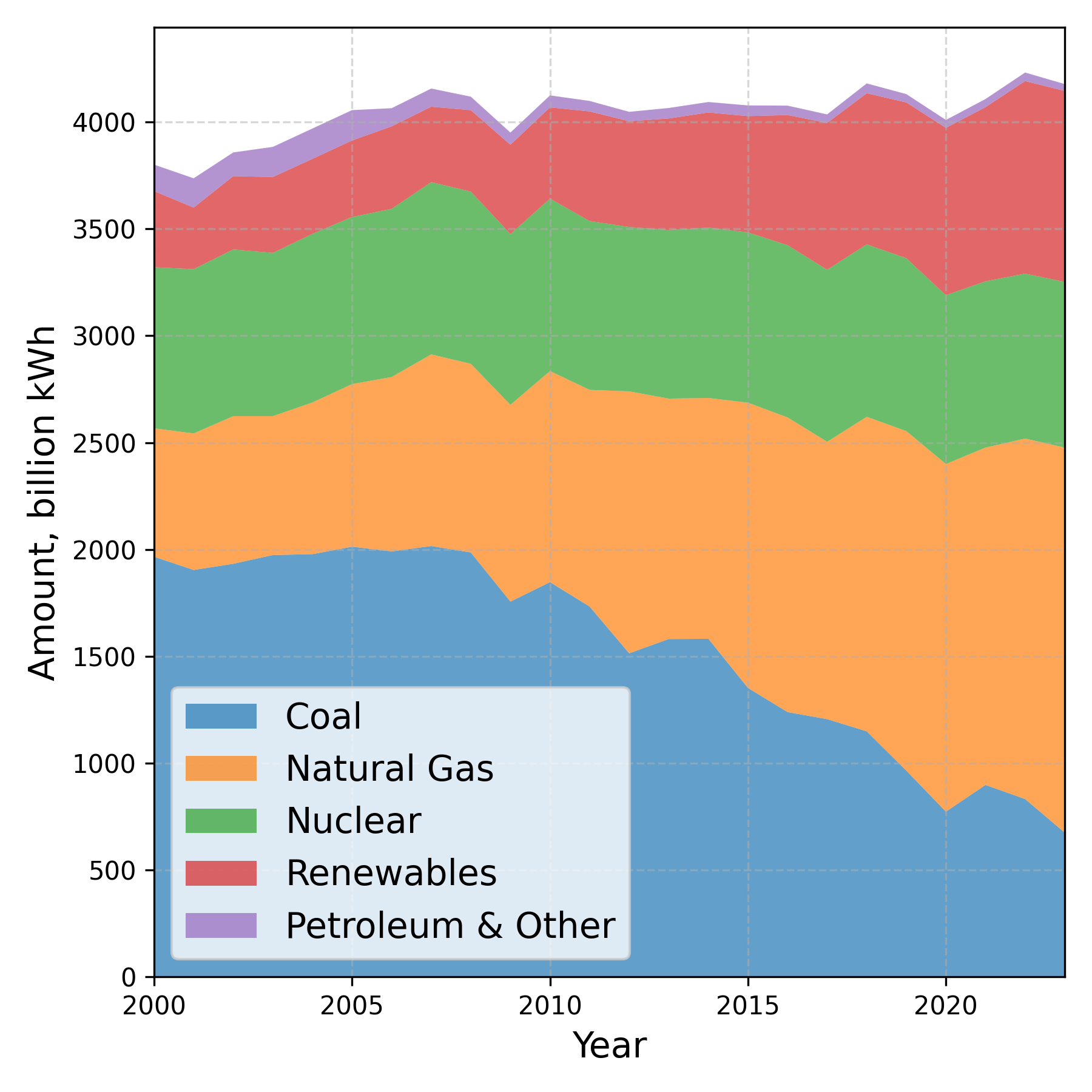}
    \caption{U.S. electricity generation mix from 2020 to 2023. Raw data sourced from the U.S. Energy Information Administration.}
    \label{fig:us-electricity-mix}
    \vspace{-5mm}
\end{wrapfigure}

In the current U.S. energy mix, the electric power sector consumes approximately one-third of total primary energy, generating nearly 4.18 trillion kWh of electricity in 2023 [25, 26]. Among various energy sources, nuclear power accounted for 9\% of total primary energy consumption, translating to 775 billion kWh of electricity, supplying 19\% of total U.S. electricity and 48\% of the nation’s carbon-free electricity [25,26,27].

In the context of the U.S. goals to achieve 100\% clean electricity and reduce greenhouse gas emissions by 61–66\% below 2005 levels by 2035, with a target of net-zero emissions by 2050\footnote{As of Jan 20, 2025, President Trump has signed an order to withdraw from the Paris Agreement, effectively nullifying emission targets set by the Biden administration. Section 1.3 is based on information from sources such as the DOE during the Biden administration and may be outdated.} [28,29], the role of nuclear power is of paramount importance, as shown in Figure \ref{fig:us-electricity-mix}.

According to the DOE’s latest report in September 2024, an additional 700–900 GW of clean, firm capacity is required to meet net-zero targets by 2050 while accommodating increasing electricity demand [30]. Given its high capacity factor, nuclear power is expected to contribute at least 200 GW of additional capacity, with DOE plans targeting 35 GW of new capacity by 2035 and a sustained growth rate of 15 GW per year by 2040 [31]. Achieving this will require continued operation and license renewal of the existing nuclear reactor fleet, as well as the construction of advanced Gen III+ and Gen IV reactors of varying scales.

The NEA has identified several sectors where SMRs could be deployed, including both electrical and non-electrical applications:
\begin{itemize}
    \item Technology Sector
    \begin{itemize}
        \item Google, Amazon, Microsoft, and Meta collectively consumed 72 TWh of electricity in 2021 [32], a figure expected to rise significantly due to increasing AI model complexity. 
        \item SMRs offer scalable and off-grid power solutions, reducing dependence on local grids while providing clean, reliable electricity for energy-intensive data centers.
        \item In October 2024, Google signed an agreement with Kairos Power to purchase electricity from its SMRs, while Amazon partnered with Energy Northwest, X-energy, and Dominion Energy on SMR development projects [33,34].
    \end{itemize}
    \item Coal-to-Nuclear Transition
    \begin{itemize}
        \item In 2022, the U.S. Energy Information Administration reported that of the 200 GW operational U.S. coal-fired capacity, a quarter is expected to retire by 2029, affecting the economies of 24 states [35].
        \item A study in 2022 on the coal-to-nuclear transition found that among a list of approximately 400 retired and operating coal plants, screened after sitting analysis, 80\% have the potential to host an advanced nuclear reactor. For a hypothetical coal plant with two 600 MW units, modeling indicates that an "all nuclear" scenario with NuScale’s 12-module VOYGR plant, could result in a net increase of over 600 jobs and a $\$$250 million in regional economic benefits per site [36].
    \end{itemize}
    \item Industry Cogeneration
    \begin{itemize}
        \item SMRs, such as high-temperature gas reactors and molten salt reactors, can provide high-temperature heat for industrial applications.
        \item When coupled with a hydrogen production plant, a 6-module NuScale SMR plant can generate 250–270 tons of hydrogen per day, supporting clean energy transition efforts [37].
    \end{itemize}
\end{itemize}

\subsection*{Reference}
\begin{enumerate}
    \item \textbf{International Atomic Energy Agency}. What are Small Modular Reactors (SMRs)? 2023.
    \item \textbf{International Atomic Energy Agency}. Power Reactor Information System SHIMANE-3. 
    \item \textbf{Hannah Ritchie}. How does the land use of different electricity sources compare? 2022.
    \item \textbf{NuScale Power}. VOYGR Power Plants.
    \item \textbf{GE Hitachi Nuclear Energy}. Small Modular Nuclear Reactors: The Future of Carbon-Free Energy. 2022.
    \item \textbf{Nuclear Energy Agency}. The NEA Small Modular Reactor Dashboard: Second Edition. 2024.
    \item \textbf{International Atomic Energy Agency}. Small Modular Reactor Advances in SMR Developments. 2024.
    \item \textbf{Nuclear Energy Agency}. Small Modular Reactors: Challenges and Opportunities. 2021.
    \item \textbf{U.S. Energy Information Administration}. Capacity outages at U.S. nuclear power plants averaged 3.1 gigawatts this summer. 2021.
    \item \textbf{International Atomic Energy Agency}. Small Modular Reactor Technology Catalogue 2024 Edition, Second edition. 2025.
    \item \textbf{U.S. Army Corps of Engineers}. Army Nuclear Power Program, 1954-1976.
    \item \textbf{U.S. Army Corps of Engineers}. Army Nuclear Power Program - Experimental Reactors.
    \item \textbf{Edwin Lyman}. The Pentagon wants to boldly go where no nuclear reactor has gone before. It won’t work. 2019.
    \item \textbf{U.S. Department of Energy}. The Story Behind America's First Potential Small Modular Reactor. 2018.
    \item \textbf{Reuters}. UPDATE 1-McDermott B\&W; unit unveils small nuclear reactor. 2009.
    \item \textbf{Nuclear Engineering International}. B\&W wins DOE funding for small modular reactor development. 2012.
    \item \textbf{American Nuclear Society}. mPower Consortium Halts Project. 2017.
    \item \textbf{U.S. Department of Energy}. NuScale Power Submits First-Ever Small Modular Reactor Design Certification Application to the Nuclear Regulatory Commission. 2017.
    \item \textbf{U.S. Department of Energy}. NRC Approves First U.S. Small Modular Reactor Design. 2020.
    \item \textbf{U.S. Department of Energy}. NRC Certifies First U.S. Small Modular Reactor Design. 2023.
    \item \textbf{Idaho National Laboratory}. Carbon Free Power Project. 
    \item \textbf{American Nuclear Society}. TerraPower submits Natrium construction application to the NRC. 2024.
    \item \textbf{American Nuclear Society}. TerraPower breaks ground on SMR project in Wyoming. 2024.
    \item \textbf{U.S. Department of Energy}. Kairos Power Starts Construction of Hermes Reactor. 2024.
    \item \textbf{U.S. Energy Information Administration}. U.S. energy facts explained. 2024.
    \item \textbf{U.S. Energy Information Administration}. What is U.S. electricity generation by energy source?. 2024.
    \item \textbf{U.S. Department of Energy}. 5 Fast Facts About Nuclear Energy. 2024.
    \item \textbf{U.S. Department of Energy}. On The Path to 100\% Clean Electricity. 2023.
    \item \textbf{United Nations Framework Convention on Climate Change}. The United States of America Nationally Determined Contribution. 2024.
    \item  \textbf{U.S. Department of Energy}. Pathways to Commercial Liftoff: Advanced Nuclear. 2024.
    \item \textbf{U.S. Department of Energy}. U.S. Sets Targets to Triple Nuclear Energy Capacity by 2050. 2024.
    \item \textbf{International Energy Agency}. Data Centres and Data Transmission Networks. 2023.
    \item \textbf{Google}. New nuclear clean energy agreement with Kairos Power. 2024.
    \item \textbf{Amazon}. Amazon signs agreements for innovative nuclear energy projects to address growing energy demands. 2024.
    \item \textbf{U.S. Energy Information Administration}. Nearly a quarter of the operating U.S. coal-fired fleet scheduled to retire by 2029. 2022.
    \item \textbf{Idaho National Laboratory}. Investigating Benefits and Challenges of Converting Retiring Coal Plants into Nuclear Plants. 2022.
    \item \textbf{NuScale Power}. The Hydrogen Rush: Ushering in a New Era of Energy with Small Modular Reactors. 2024.
\end{enumerate}

\clearpage
\newpage
\section{Notable SMR Designs}
\subsection{AP300 (Westinghouse Electric Company)}

\begin{wraptable}{o}{9cm} 
    \vspace{-12mm} 
    \centering
    \begin{threeparttable} 
        \renewcommand{\arraystretch}{1.2} 
        \rowcolors{2}{gray!15}{white} 
        \begin{tabular}{ p{4cm} | p{4cm} } 
            \toprule
            \rowcolor{white}
            \multicolumn{2}{l}{\textbf{General Information}} \\ 
            \midrule
            Reactor Type & Pressurized Water Reactor \\
            Purpose & Commercial \\
            Thermal Power (MWt) & 990 \\
            Net Power Output (MWe) & 330 \\
            Design Life & 80 years \\
            Reactor Units per Site & Single unit \\
            Seismic Design & 0.3 g \\ 
            Site Footprint (m\textsuperscript{2}) & 58,000 \\
            Construction Time (NOAK) & 36 months \\
            
            \midrule
            \rowcolor{white}
            \multicolumn{2}{l}{\textbf{Fuel $\&$ Materials}} \\ 
            \midrule
            
            Core Coolant & H\textsubscript{2}O \\
            Neutron Moderator & H\textsubscript{2}O \\
            Solid Burnable Absorber  & Gd\textsubscript{2}O\textsubscript{3} \\
            Fuel Cladding & Zr-Alloy \\
            Fuel Material & UO\textsubscript{2} \\
            Fuel Enrichment & Less than 5\% \\
            Refuelling Cycle & 36 to 48 months\\
            
            \midrule
            \rowcolor{white}
            \multicolumn{2}{l}{\textbf{Development $\&$ Licensing}} \\ 
            \midrule
            
            Design Status & Basic Design\\
            Licensing Status &  - \\
            \bottomrule
        \end{tabular}
        \begin{tablenotes}
          \item {Last ARIS update on 2024/10/10}
        \end{tablenotes}
    \end{threeparttable}
    \vspace{-12mm} 
\end{wraptable}

Due to the lack of publicly available Topical Reports or White Papers on the AP300 at the time of writing, most of the technical information presented here is inferred from the AP1000 design.

\paragraph{Basic Design}
The AP300 is a 330 MWe pressurized water small modular reactor designed by Westinghouse, based on the proven technology of the AP1000 reactor [1]. The AP300 retains key design elements from the AP1000, including reactor and fuel design as well as passive safety features. Six units of the AP1000 reactor are currently in operation in China and the U.S., offering valuable operating experience to the development of the AP300.

Compared to the AP1000, the AP300 features a simplified Nuclear Steam Supply System with a one-loop configuration, comprising one steam generator, one hot leg, and two cold legs, each equipped with a reactor coolant pump. The core consists of 121 fuel assemblies (down from 157 in the AP1000) in a 17$\times$17 Robust Fuel Assembly configuration, with 264 fuel rods per assembly and a refueling period of up to four years [1, 2].

\paragraph{Fuel Design}
Since the AP300 shares the AP1000’s fuel design, it is expected to use enriched uranium dioxide fuel pellets (<5 wt.\% U-235) with integrated burnable absorbers, either as a boron coating (<0.001 inch thick) or Gd\textsubscript{2}O\textsubscript{3} dopants [2, 3, 4]. Reports suggest that the AP300 incorporates Westinghouse’s Advanced Doped Pellet Technology fuel, which contains small amounts of Cr\textsubscript{2}O\textsubscript{3} and Al\textsubscript{2}O\textsubscript{3} to improve fuel performance and stability; however, official confirmation is lacking [5,6]. Unlike the AP1000, where the spent fuel pool is inside the auxiliary building, the AP300 places it inside the containment building [7].

\paragraph{Reactivity Control}
The AP300 employs control rods, gray rods, and the Mechanical Shim strategy from the AP1000 to simultaneously regulate reactivity and power distribution [7,8]. While there is limited information on the AP300, based on AP1000 documentation, it is likely to use silver-indium-cadmium control rod clusters for shutdown and rapid reactivity changes. The gray rod clusters, consisting of half stainless steel rods and half reduced-diameter Ag-In-Cd rods clad in stainless steel, offer more precise reactivity control, reducing the need for soluble boron adjustments [3].

\paragraph{Safety Features}
The AP300’s safety system is designed to be passive, operating without active components such as pumps or diesel generators, and without external power sources such as AC electricity, leveraging experience from the AP1000’s design. It also minimizes the need for operator intervention to achieve safe shutdown under all design-basis events [8].

In the event of a loss-of-coolant accident or main steam line break, the AP1000’s Passive Containment Cooling System uses the passive containment cooling water storage tank located above the containment to cool its outer stainless steel surface, while natural air circulation further facilitates heat removal. The Passive Core Cooling System, including the Core Makeup Tanks, Accumulators, and the In-Containment Refueling Water Storage Tank, activate at different stages to inject borated water into the reactor vessel in response to depressurization. The In-Containment Refueling Water Storage Tank also plays a role in long-term decay heat removal via the Passive Residual Heat Removal Heat Exchanger [9].

In the AP1000 design, these passive safety systems provide a 72-hour coping time, during which no operator action is required. After this period, operators can replenish water from auxiliary storage tanks and other sources to sustain core and containment cooling indefinitely [10].

\paragraph{Development Timeline}
On May 4, 2023, Westinghouse officially announced the AP300 SMR, a scaled-down version of the AP1000 reactor [11]. Soon after, on May 9, Westinghouse submitted a Regulatory Engagement Plan to the NRC, outlining its approach to licensing the AP300 SMR [12]. The company has since committed to periodic updates, with the latest Revision C submitted in November 2024, detailing advancements and outlining future pre-application interactions with the NRC [7].

In the second quarter of 2024, Westinghouse began preparing the AP300 SMR Design Control Document, leveraging existing AP1000 documentation. According to the AP300 SMR U.S. Licensing Roadmap, Westinghouse aims to obtain Design Certification from the NRC by 2027, with standard AP300 plants ready for deployment [7].

Westinghouse is actively engaged in the overseas market, announcing plans to deploy four AP300 SMRs in Northeast England by the early 2030s [13]. In August 2024, it received approval from the UK Department of Energy Security and Net Zero to enter the Generic Design Assessment, the first step in the UK’s licensing process [14].

\subsection*{Reference}
\begin{enumerate}
    \item\textbf{International Atomic Energy Agency}. Small Modular Reactor Technology Catalogue 2024 Edition, Second edition. 2025. 
    \item \textbf{Westinghouse Electric Company}. Westinghouse AP1000 Design Control Document Rev. 19 - Tier 2 Chapter 4 – Reactor – Section 4.1 Summary Description. 2011.
    \item \textbf{Westinghouse Electric Company}.Westinghouse AP1000 Design Control Document Rev. 19 - Tier 2 Chapter 4 – Reactor – Section 4.2 Fuel System Design. 2011.
    \item \textbf{Westinghouse Electric Company}.Westinghouse AP1000 Design Control Document Rev. 19 - Tier 2 Chapter 4 – Reactor – Section 4.3 Nuclear Design. 2011.
    \item \textbf{Power Magazine}. Westinghouse Unveils the AP300—A Miniaturized AP1000 Small Modular Nuclear Reactor. 2023.
    \item \textbf{Westinghouse Electric Company}. ADOPT Fuel.
    \item \textbf{Westinghouse Electric Company}. Westinghouse AP300 SMR Pre-Application Regulatory Engagement Plan Periodic Update – November 2024. 2024.
    \item \textbf{Westinghouse Electric Company}. Transmittal of Public Session Presentation Slides to Support the Westinghouse-NRC Pre-Submittal Meeting on the Westinghouse AP300 SMR Core Design \& Fuel Cycle White Paper. 2024.
    \item \textbf{Westinghouse Electric Company}. AP1000 The PWR Revisited.
    \item \textbf{Westinghouse Electric Company}. AP1000 Plant Passive Safety Systems and Timeline for Station Blackout. 2021.
    \item \textbf{Westinghouse Electric Company}. Westinghouse Unveils Game-Changing AP300 Small Modular Reactor for Mid-Sized Nuclear Technology. 2023.
    \item \textbf{Westinghouse Electric Company}. Westinghouse Submits AP300 SMR Regulatory Engagement Plan to Nuclear Regulatory Commission. 2023.
    \item \textbf{Westinghouse Electric Company}. Westinghouse and UK’s Community Nuclear Power Collaborate to Deploy Fleet of AP300 Small Modular Reactors. 2024.
    \item \textbf{Westinghouse Electric Company}. Westinghouse AP300 Small Modular Reactor Approved for United Kingdom’s Generic Design Assessment. 2024.
\end{enumerate}

\clearpage
\newpage
\subsection{NuScale Power Module (NuScale Power)}

\begin{wraptable}{r}{9cm} 
    \vspace{-12mm} 
    \centering
    \begin{threeparttable} 
        \renewcommand{\arraystretch}{1.2} 
        \rowcolors{2}{gray!15}{white} 
        \begin{tabular}{ p{4cm} | p{4cm} } 
            \toprule
            \rowcolor{white}
            \multicolumn{2}{l}{\textbf{General Information}} \\ 
            \midrule
            Reactor Type & Pressurized Water Reactor \\
            Purpose & Commercial \\
            Thermal Power (MWt) & 250 \\
            Gross Power Output (MWe) & 77 \\
            Design Life & 60 years \\
            Reactor Units per Site & - \\
            Seismic Design & 0.5 g \\ 
            Site Footprint (m\textsuperscript{2}) & - \\
            Construction Time (NOAK) & - \\
            
            \midrule
            \rowcolor{white}
            \multicolumn{2}{l}{\textbf{Fuel $\&$ Materials}} \\ 
            \midrule
            
            Core Coolant & H\textsubscript{2}O \\
            Neutron Moderator & H\textsubscript{2}O \\
            Solid Burnable Absorber  & - \\
            Fuel Cladding & - \\
            Fuel Material & UO\textsubscript{2} \\
            Fuel Enrichment & Less than 4.95\% \\
            Refueling Cycle & 18 months \\
            
            \midrule
            \rowcolor{white}
            \multicolumn{2}{l}{\textbf{Development $\&$ Licensing}} \\ 
            \midrule
            
            Design Status & Under Regulatory Review\\
            Licensing Status & -   
            \end{tabular}

        \begin{tablenotes}
           \item {Last ARIS update on 2024/10/14}
        \end{tablenotes}
    \end{threeparttable}
    \vspace{-10mm} 
\end{wraptable}

Considering that NuScale Power has submitted its Standard Design Approval application to the NRC, the majority of the technical details presented here are derived from the application materials.

\paragraph{Basic Design}
The NuScale advanced SMR is a scalable design, accommodating one to six NuScale Power Modules, each with a thermal output of 250 MWt and an electrical generation capacity of 77 MWe, enabling a six-module plant to produce 462 MWe. The design leverages existing PWR technology, with a fuel assembly similar to conventional PWRs [1].

\begin{figure}
    \centering
    \includegraphics[width=\linewidth]{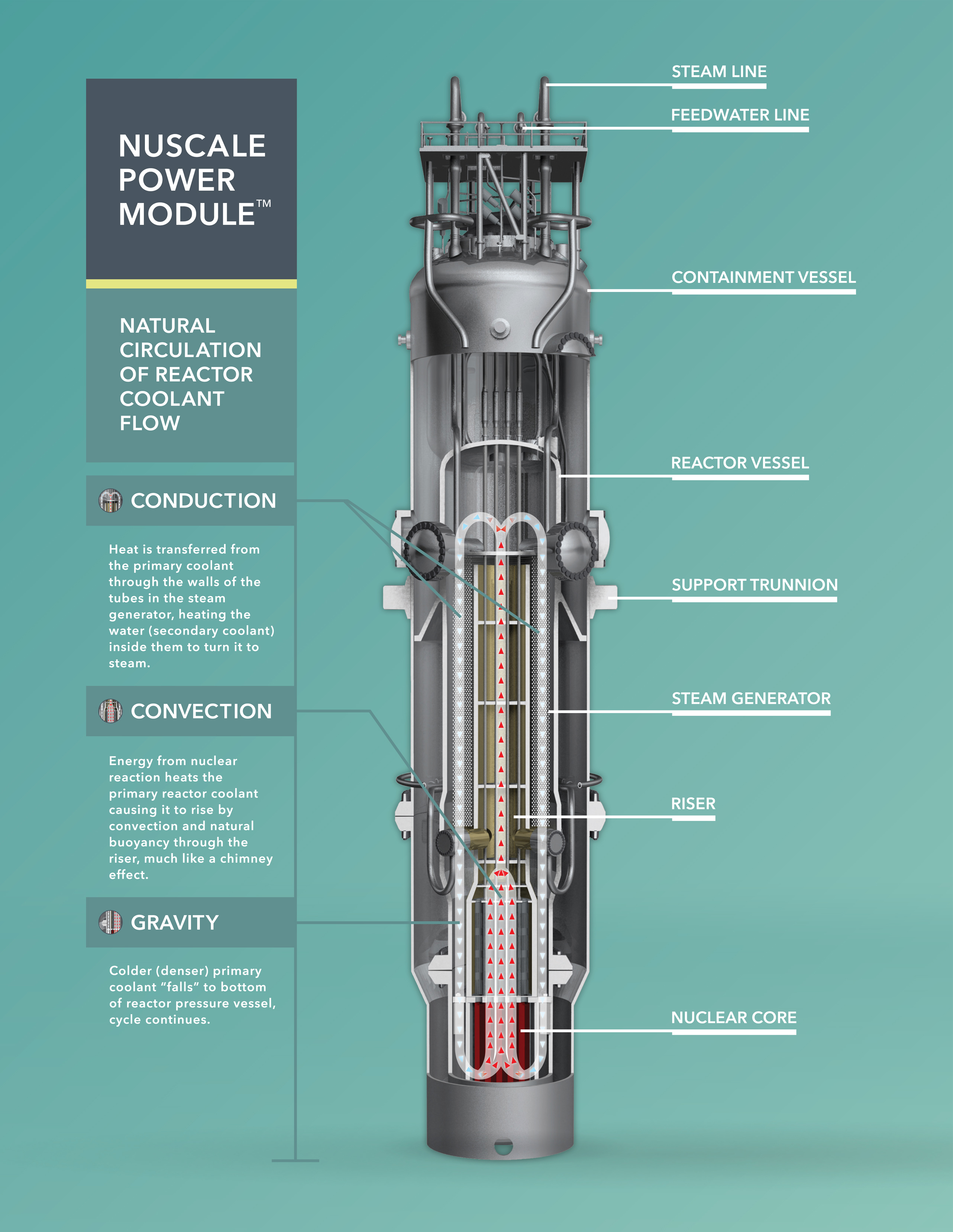}
    \caption{Diagram of a NuScale Power Module. Image Credit: NuScale Power}
    \label{fig:nu-scale}
\end{figure}

Each Power Module integrates a reactor core, pressurizer, and two steam generators within a pressure vessel, enclosed in a compact cylindrical steel containment vessel, as illustrated in Figure \ref{fig:nu-scale}. The containment vessel is partially submerged in water, providing a passive heat sink for long-term decay heat removal. The design eliminates reactor coolant pumps, instead relying on natural circulation for reactor cooling. During operation, heated coolant rises through the upper riser assembly, transfers heat to the steam generators and returns to the bottom of the pressure vessel, maintaining continuous reactor coolant flow [1].

The design also eliminates reliance on external power for safe shutdown and cooling, instead utilizing module-specific passive safety systems and an Ultimate Heat Sink that includes the reactor pool, refueling pool, and spent fuel pool. This configuration allows individual Power Module to be installed, tested, and started up independently without affecting operating units [1].

The six-module plant layout consists of a reactor building (housing Power Modules and operational systems) and a control building, where operators monitor and control all installed Power Modules [1]. The main control room requires at least one licensed reactor operator and two licensed senior reactor operators per shift to maintain plant operations [2]. 

\paragraph{Fuel Design}
The Power Module uses a 17×17 PWR fuel assembly, the NuFuel-HTP2, developed based on the existing Framatome HTP design. Each reactor contains 37 fuel assemblies, with 264 fuel rods, 24 guide rods, and a central instrument tube per assembly. Due to the Power Module's compact core, the NuFuel-HTP2 is approximately half the length of standard PWR fuel assemblies [3,4].

Each fuel rod consists of sintered UO\textsubscript{2} pellets with a theoretical density of 96.5\%, enriched up to 4.95 wt\% U-235, housed within M5 zirconium alloy cladding and filled with helium gas. The rod void volume is designed to accommodate fission gas generation, ensuring that internal pressure remains below reactor coolant system pressure. Additionally, the design allows for axial blanket and Gd\textsubscript{2}O\textsubscript{3} fuel configurations, both of which share the same structural geometry as standard UO\textsubscript{2} fuel rods [3,4]. 

Refueling occurs every 18 months, with new and spent fuel assemblies stored in the spent fuel pool located in the reactor building [4,5]. Each refueling operation is independent of other Power Modules, as the fuel handling system moves only one assembly at a time, allowing individual module refueling without disrupting other units. The spent fuel pool has a maximum capacity of 600 fuel assemblies and is actively cooled under normal conditions. During accident scenarios, the Ultimate Heat Sink provides passive cooling to ensure fuel safety [5].

\clearpage
\newpage
\paragraph{Reactivity Control}
The Power Module employs three primary methods for reactivity control: soluble boron, control rod assemblies, and burnable poison. Soluble boron and control rods are used for power maneuvering, allowing fine-tuned adjustments to reactor power levels. In contrast, excess reactivity throughout the fuel cycle is managed by burnable poisons and soluble boron, and it is not a requirement for the control rod assemblies [3].

Soluble boron is used during operation to compensate for reactivity changes due to power level, fuel burnup, fission product poisoning, and burnable poison depletion [3]. Its concentration is adjusted throughout the cycle, with higher initial levels balancing the excess reactivity at startup. During normal operations, soluble boron is introduced into the reactor coolant system via the Chemical and Volume Control System and the Boron Addition System [5]. Additionally, upon actuation of the Emergency Core Cooling System during emergency scenarios, the Emergency Core Cooling System supplementary boron system injects additional soluble boron to ensure the reactor remains subcritical [1].

To reduce the need for high boron concentrations early in the cycle, NuScale incorporates Gd\textsubscript{2}O\textsubscript{3} burnable poison within selected fuel rods by homogeneously mixing it with UO\textsubscript{2} [3,4]. During the fuel cycle, Gd\textsubscript{2}O\textsubscript{3} helps control radial power distribution and limit excessive local reactivity.

Among the 37 fuel assemblies in a Power Module, 16 positions contain control rod assemblies. Each control rod assembly consists of 24 individual control rods, with boron carbide pellets in the upper section and silver-indium-cadmium absorber material in the lower section. The rods are enclosed in stainless steel cladding and filled with helium gas [3,4]. NuScale’s control rod assemblies are based on Framatome’s Rod Control Cluster Assembly design, retaining the basic features and materials. However, due to Power Module's smaller core size, the control rods, like the fuel rods, are shorter than those used in conventional reactors [4].

The control rod assemblies are divided into eight regulating banks and eight shutdown banks [3]. The regulating banks are split into two symmetrical groups, which adjust reactivity and shape the axial power distribution during normal operations. The shutdown banks, also split into two groups, remain fully withdrawn during operation but are inserted during reactor trips to maintain shutdown conditions\footnote{Figure 4.3-14 in the Final Safety Analysis Report shows the shutdown bank as four groups of two.}. During startup, the shutdown bank is fully withdrawn before the regulating bank is slowly removed, in conjunction with boron dilution, to approach criticality [3].

\paragraph{Safety Features}
The Power Module employs a passive safety design that significantly reduces the number of required safety components compared to conventional PWRs. The pressure vessel integrates the core, steam generators, and pressurizer within a single structure, eliminating external piping and minimizing the risk of large-break LOCAs [1]. Acting as a primary containment barrier, the pressure vessel prevents fission product release while supporting key safety mechanisms, including control rod drive mechanisms, reactor safety valves, and Emergency Core Cooling System valves [6]. 

The Emergency Core Cooling System is a fully passive, safety-related system responsible for heat removal, without the need for external power. It ensures decay and residual heat removal when reactor coolant inventory redistributes between the pressure vessel and the containment vessel due to events such as a pipe break LOCA [7,8]. The Emergency Core Cooling System consists of two sets of valves: the Reactor Vent Valves and Reactor Recirculation Valves, which remain closed during normal operation but automatically open upon Emergency Core Cooling System actuation. Once triggered, steam is vented from the pressure vessel into the containment vessel, where it condenses on the containment walls, forming a liquid coolant pool at the bottom. The containment vessel is partially submerged in the Ultimate Heat Sink, which provides immediate and continuous heat absorption. As the pressures of the pressure vessel and containment vessel equalize, the coolant passively circulates back into the pressure vessel through the Reactor Recirculation Valves, sustaining core cooling. The placement of the Reactor Recirculation Valve penetrations ensures that coolant levels remain above the core, preventing fuel exposure and maintaining effective decay heat removal [7,8].

In addition to the Emergency Core Cooling System, the passive, safety-related Decay Heat Removal System serves as a backup cooling mechanism when normal secondary-side cooling is unavailable [6]. Each steam generator is connected to a Decay Heat Removal System train, which consists of passive condensers mounted on the outer surface of the containment. These condensers, submerged in the Ultimate Heat Source, utilize natural circulation to facilitate coolant flow. Positioned above the steam generators, the Decay Heat Removal System condensers establish a two-phase cooling loop, leveraging density variations between steam and condensate to return cooled fluid to the steam generators without requiring active pumps [6,8].

Finally, the Ultimate Heat Sink is a critical component of the passive heat removal systems and includes redundant water level instrumentation and a dedicated makeup line to extend cooling if necessary [5]. Under normal operating conditions, heat is dissipated from the heat sink through a cooling system and ultimately released into the atmosphere via a cooling tower or another external heat sink. In the event of power loss, reactor heat is passively removed by allowing the reactor pool to heat up and boil. The reactor pool's water inventory is maintained at a level sufficient to enable safety-related systems to sustain decay and residual heat removal for a minimum of 72 hours without operator intervention during design-basis events [8].

\paragraph{Development Timeline}
In 2013, NuScale Power was selected as the winner of the second round of the DOE’s cost-sharing program for SMR development, securing up to \$226 million over five years [9]. This funding was formalized in May 2014, when NuScale and the DOE signed a cooperative agreement, initiating up to \$217 million in matching funds for SMR development [10]. Additionally, in 2018, NuScale received \$80 million under the First-of-a-Kind Nuclear Demonstration Readiness Project to support the commercial operation of the first NuScale plant by 2026 [11].

In January 2017, NuScale submitted its Design Certification application to the NRC, receiving approval in 2020, and officially becoming the first SMR design certified by the NRC in 2023 [12,13]. This certification applied to the NuScale SMR US600, which featured 12 modules, each rated at 50 MWe.

NuScale was also involved in the Carbon Free Power Project, launched in 2015 by the Utah Associated Municipal Power Systems to deploy a NuScale SMR at Idaho National Laboratory with operations originally targeted for 2030 [14]. The project received up to \$1.4 billion in DOE cost-share funding in 2020 [15]. However, in 2021, the Utah Associated Municipal Power Systems revised the CFPP, reducing the plant to six modules and 462 MWe, with each module rated at 77 MWe [16]. In response, NuScale submitted a Standard Design Approval application in 2023 for the revised US460 model, which is expected to complete NRC review by 2025 [17]. Despite these efforts, on November 8, 2023, the Utah Associated Municipal Power Systems and NuScale mutually agreed to terminate the CFPP [14].

Internationally, NuScale Power has pursued licensing and deployment agreements across multiple countries. In December 2019, NuScale initiated a Vendor Design Review process with the Canadian Nuclear Safety Commission [18], though no formal progress records are currently available. As of 2023, NuScale Power has 19 signed and active agreements to deploy its SMR over 12 countries, according to the DOE [13].

\paragraph{Reference}
\begin{enumerate}
    \item \textbf{NuScale Power}. NuScale US460 Plant Standard Design Approval Application, Chapter One, Introduction and General Description of the Plan. 2023.
    \item \textbf{NuScale Power}. NuScale US460 Plant Standard Design Approval Application, Chapter Eighteen, Human Factors Engineering. 2023.
    \item \textbf{NuScale Power}. NuScale US460 Plant Standard Design Approval Application, Chapter Four, Reactor. 2023.
    \item \textbf{NuScale Power}. Licensing Technical Report, NuFuel-HTP2 Fuel and Control Rod Assembly Designs. 2022.
    \item \textbf{NuScale Power}. NuScale US460 Plant Standard Design Approval Application, Chapter Nine, Auxiliary Systems. 2023.
    \item \textbf{NuScale Power}. NuScale US460 Plant Standard Design Approval Application, Chapter Five, Reactor Coolant System and Connecting Systems. 2023.
    \item \textbf{NuScale Power}. NuScale US460 Plant Standard Design Approval Application, Chapter Six, Engineered Safety Features. 2023.
    \item \textbf{NuScale Power}. Licensing Topical Report, Extended Passive Cooling and Reactivity Control Methodology. 2022.
    \item \textbf{NuScale Power}. U.S. DOE Awards Funding for NuScale Power’s SMR Technology. 2013.
    \item \textbf{NuScale Power}. NuScale and DOE Complete SMR Cooperative Agreement. 2014.
    \item \textbf{U.S. Department of Energy}. Secretary of Energy Rick Perry Announces \$60 Million for U.S. Industry Awards in Support of Advanced Nuclear Technology Development. 2018.
    \item \textbf{U.S. Nuclear Regulatory Commission}. Design Certification - NuScale US600.
    \item \textbf{U.S. Department of Energy}. NRC Certifies First U.S. Small Modular Reactor Design. 2023.
    \item \textbf{Idaho National Laboratory}. Carbon Free Power Project.
    \item \textbf{U.S. Department of Energy}. DOE Approves Award for Carbon Free Power Project. 2020.
    \item \textbf{American Nuclear Soceity}. UAMPS downsizes NuScale SMR plans. 2021.
    \item \textbf{U.S. Nuclear Regulatory Commission}. NuScale US460 Standard Design Approval Application Review.
    \item \textbf{Business Wire}. NuScale Submits Phase 1 and 2 Combined Pre-Licensing Vendor Design Review to Canadian Nuclear Safety Commission. 2020.
\end{enumerate}

\clearpage
\newpage
\subsection{PWR-20 (Last Energy)}

\begin{wraptable}{o}{9.5cm} 
    \vspace{-12mm} 
    \centering
    \begin{threeparttable} 
        \renewcommand{\arraystretch}{1.2} 
        \rowcolors{2}{gray!15}{white} 
        \begin{tabular}{ p{4cm} | p{4.5cm} } 
            \toprule
            \rowcolor{white}
            \multicolumn{2}{l}{\textbf{General Information}} \\ 
            \midrule
            Reactor Type & Pressurized Water Reactor \\
            Purpose & Commercial \\
            Thermal Power (MWt) & 80 \\
            Net Power Output (MWe) & 20 \\
            Design Life & 42 years \\
            Reactor Units per Site & Multiple units \\
            Seismic Design & 0.5 g horizontal; 0.4 g vertical \\ 
            Site Footprint (m\textsuperscript{2}) & 17,500  \\
            Construction Time (NOAK) & 6 \\
            
            \midrule
            \rowcolor{white}
            \multicolumn{2}{l}{\textbf{Fuel $\&$ Materials}} \\ 
            \midrule
            
            Core Coolant & H\textsubscript{2}O \\
            Neutron Moderator & H\textsubscript{2}O \\
            Solid Burnable Absorber  & Gd\textsubscript{2}O\textsubscript{3} \\
            Fuel Cladding & Zr-4 \\
            Fuel Material & UO\textsubscript{2} \\
            Fuel Enrichment & Less than 4.95\% \\
            Refueling Cycle & 72 months \\
            
            \midrule
            \rowcolor{white}
            \multicolumn{2}{l}{\textbf{Development $\&$ Licensing}} \\ 
            \midrule
            
            Design Status & Detailed Design \\
            Licensing Status & -   
            \end{tabular}
            
        \begin{tablenotes}
            \item {Last ARIS update on 2024/10/15}
        \end{tablenotes}    
    \end{threeparttable}
    \vspace{-10mm} 
\end{wraptable}

Compared to other advanced LWRs such as the AP300, NuScale Power Modules, and BWRX-300, the PWR-20 has no publicly available licensing documents discussing its design docketed by the NRC as of March 2025. Information about its design is primarily sourced from Last Energy’s own materials and the IAEA.

\paragraph{Basic Design}
The PWR-20 is a single-loop pressurized water reactor\footnote{Occasionally referred to as a (scaled-down) four-loop PWR by Last Energy} designed to generate 20 MWe and continous thermal energy at 300\textsuperscript{o}C for industrial applications [1,2].

The design leverages conventional PWR technology but focuses on deployment innovation, with Last Energy claiming a 24-month delivery timeline from reactor fabrication to on-site assembly completion [3]. This rapid deployment is attributed to the use of off-the-shelf equipment, such as steam turbine generators, heat exchangers, and steam generators [4].

\begin{figure}[h]
    \centering
    \includegraphics[width=0.9\linewidth]{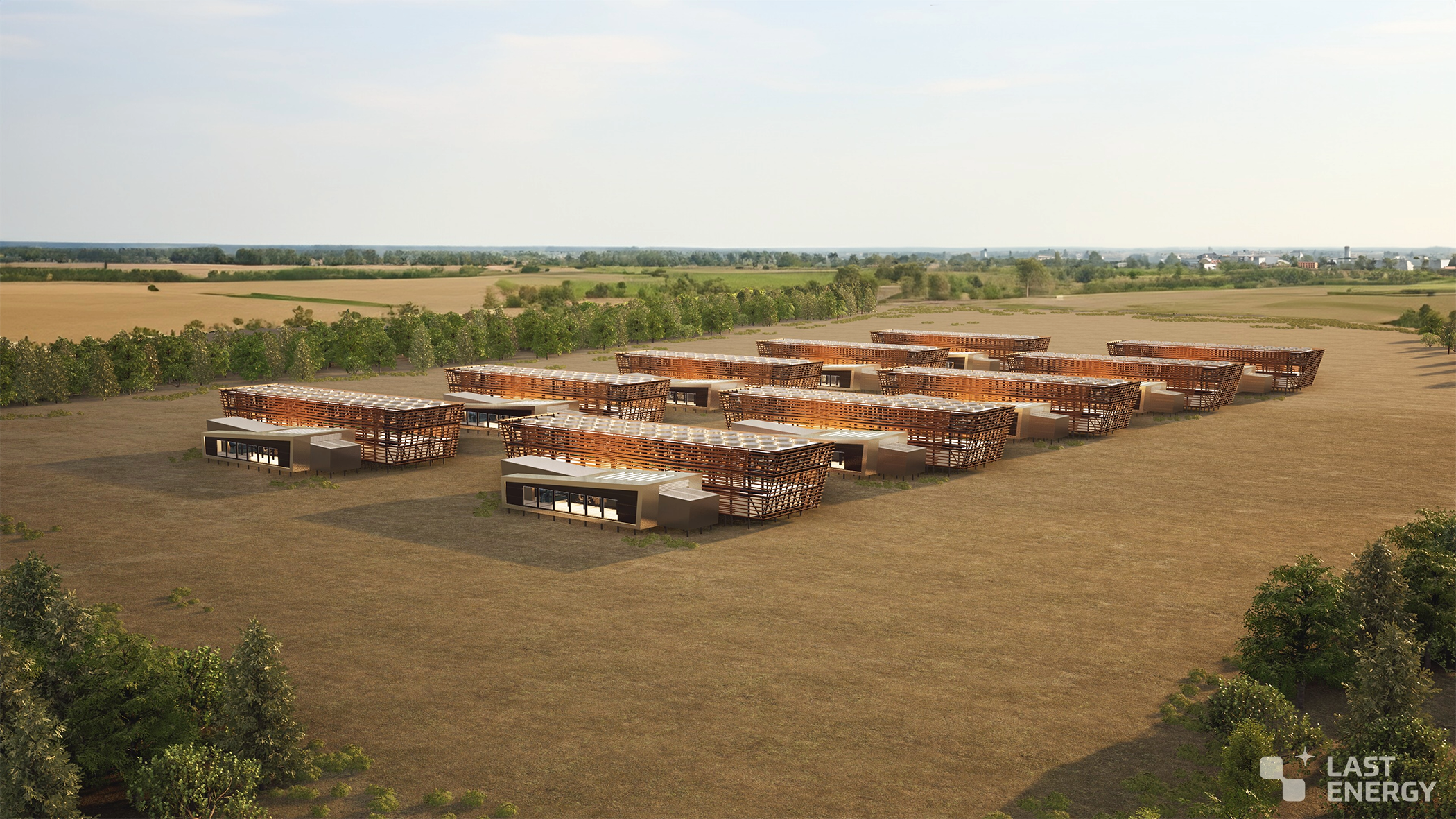}
    \caption{Rendered site layout of a ten-unit PWR-20 reactor facility. Image credit: Last Energy}
    \label{fig:10-unit-pwr}
\end{figure}

Each PWR-20 reactor plant is designed to be compact, occupying less than 0.5 acres, making it well-suited for industrial sites [5]. Figure \ref{fig:10-unit-pwr} illustrates a site layout featuring ten PWR-20 reactor plants, showcasing how multiple units can be arranged within a small footprint. Additionally, the PWR-20’s tertiary loop is air-cooled, significantly reducing water consumption to less than 1 gallon per minute and eliminating the need for a nearby water source. As a result, Last Energy describes the PWR-20 as having "near-universal siting" potential [2]. 

\paragraph{Fuel Design}
The PWR-20 reactor core and fuel design follow conventional PWR industry standards, utilizing a 17$\times$17 fuel assembly configuration with low-enriched uranium dioxide (UO\textsubscript{2}) fuel enriched below 4.95 wt.\% [1,2]. According to the IAEA, reactivity control is achieved using both control rods and burnable poisons [1].

The reactor operates on a 72-month fuel cycle, followed by a 3-month refueling period [1]. However, neither the IAEA report nor Last Energy provides detailed information on the refueling process or spent fuel management. 

Last Energy describes the refueling process as replacing the spent fuel module with a new, pre-fueled module installed directly into the power plant. The removed reactor module is left to cool for the remainder of the plant’s operational life, with both on-site wet and dry fuel storage available prior to eventual off-site transfer [6].

\paragraph{Safety Features}
The PWR-20 safety system is designed to be fully passive, eliminating the need for active operator control or external power sources in emergencies. The containment system consists of a 500-metric-ton iron cask, preventing the release of radioactive material in accident scenarios. PWR-20 is designed with a core damage frequency of 10\textsuperscript{-7} per reactor-year, ensuring that no core melt scenarios can occur [1].

However, beyond the IAEA report, there is limited publicly available information detailing PWR-20’s safety mechanisms.

\paragraph{Development Timeline}
Last Energy was launched in 2019 by the Energy Impact Center, focusing on the development of commercial small modular nuclear power plants. The company has made notable progress in securing commercial agreements, including Power Purchase Agreements for 34 reactors with industrial partners in Poland and the UK in 2023 [7]. By mid-2024, reports indicated that Last Energy had signed commercial agreements for over 80 units across Europe, with nearly half allocated to data centers [8].

Despite ongoing licensing efforts, there are no publicly available records confirming that Last Energy has received formal regulatory approval in either Poland or the UK as of March 2025. Back in 2024, Last Energy completed the pre-authorization phase of the licensing process in Romania and has been allowed to enter the authorization phase, but full approval has not yet been granted [9]. Meanwhile, in the United States, Last Energy is engaged in pre-application activities regarding an Early Site Permit planned for submission in June 2025, according to the NRC. As of March 2025, no topical reports, white papers, or meeting materials on the PWR-20 have been made publicly available.

\paragraph{Reference}
\begin{enumerate}
    \item \textbf{International Atomic Energy Agency}. Small Modular Reactor Technology Catalogue 2024 Edition, Second edition. 2025.
    \item \textbf{Last Energy}. General Fact Sheet. 
    \item \textbf{Last Energy}. General Brochure.
    \item \textbf{Last Energy}. Standardization to Scale Nuclear. 2023.
    \item \textbf{Last Energy}. Siting Fact Sheet.
    \item \textbf{Last Energy}. FAQ: Plant Construction and Operations.
    \item \textbf{American Nuclear Society}. Last Energy sets up microreactor deals for Poland and the U.K. 2023.    
    \item \textbf{World Nuclear News}. Funding and commercial agreements grow for Last Energy. 2024.
    \item \textbf{Last Energy}. Last Energy Enters Authorization Phase With Nuclear Regulator in Romania. 2024.
\end{enumerate}

\clearpage
\newpage
\subsection{SMR-300 (Holtec International)}

\begin{wraptable}{o}{9cm} 
    \vspace{-12mm} 
    \centering
    \begin{threeparttable} 
        \renewcommand{\arraystretch}{1.2} 
        \rowcolors{2}{gray!15}{white} 
        \begin{tabular}{ p{4cm} | p{4cm} } 
            \toprule
            \rowcolor{white}
            \multicolumn{2}{l}{\textbf{General Information}} \\ 
            \midrule
            Reactor Type & Pressurized Water Reactor \\
            Purpose & Prototype \\
            Thermal Power (MWt) & 1000 \\
            Net Power Output (MWe) & 300 \\
            Design Life & 80 years \\
            Reactor Units per Site & Dual unit \\
            Seismic Design & 0.4 g \\ 
            Site Footprint (m\textsuperscript{2}) & 80,000 for dual units\\
            Construction Time (NOAK) & 24 months \\
            
            \midrule
            \rowcolor{white}
            \multicolumn{2}{l}{\textbf{Fuel $\&$ Materials}} \\ 
            \midrule
            
            Core Coolant & H\textsubscript{2}O \\
            Neutron Moderator & H\textsubscript{2}O \\
            Solid Burnable Absorber  & Gd\textsubscript{2}O\textsubscript{3} \\
            Fuel Cladding & Zr-4 \\
            Fuel Material & UO\textsubscript{2} \\
            Fuel Enrichment & avg 4.9\%, max 5\% \\
            Refueling Cycle & 18 months \\
            
            \midrule
            \rowcolor{white}
            \multicolumn{2}{l}{\textbf{Development $\&$ Licensing}} \\ 
            \midrule
            
            Design Status & Detailed Design \\
            Licensing Status & Preliminary Safety Analysis Report   
            \end{tabular}

        \begin{tablenotes}
            \item {Last ARIS update on 2024/06/04}
        \end{tablenotes}
    \end{threeparttable}
    \vspace{-8mm} 
\end{wraptable}

Holtec’s Small Modular Reactor (SMR-300) is an evolution of its earlier SMR-160 design. Pre-application activities for SMR-160 with the NRC were suspended in late 2023, shifting focus to SMR-300 [1]. Due to the limited publicly available documentation on SMR-300, this summary integrates design information from both designs.

\paragraph{Basic Design}
Designed primarily for electricity generation, the SMR-300 also supports various cogeneration applications, such as hydrogen production and district heating. SMR-300 has a designed thermal capacity of 1050 MW and produces approximately 300 MWe, with an expected operational lifespan of 80 years [2]. It employs a forced circulation system with two reactor coolant pumps and a dual-loop configuration consisting of two hot legs and two cold legs. Heat transfer occurs through a once-through steam generator, with the pressurizer positioned directly above to optimize efficiency [2].

The design prioritizes passive safety, relying exclusively on natural forces for core cooling without the need for external power sources, pumps, or operator intervention. This eliminates dependence on offsite infrastructure, enhancing resilience in emergency scenarios. Additionally, unlike conventional nuclear power plants that require substantial water resources, the SMR-300 can be configured with an air-cooling system, allowing it to operate in water-scarce environments [3].

\paragraph{Fuel Design}
The SMR-300 reactor core consists of 69 uranium dioxide fuel assemblies arranged in a 17$\times$17 configuration, with a maximum uranium enrichment level of 5\% [4]. The reactor follows an 18-month refueling cycle and utilizes Framatome’s GAIA 17x17 fuel assemblies along with the HARMONI rod cluster control assembly, components already in use in existing pressurized water reactors, ensuring proven performance [4]. 

SMR-300's spent fuel pool is located adjacent to the reactor vessel inside the containment, facilitating a streamlined refueling process that is projected to take no more than seven days. The pool typically hosts only a single batch of recently discharged fuel at any given time, leading to a much smaller fissile material inventory compared to a standard pressurized water reactor [3]. In the case of SMR-160, a refueling water storage tank supplies borated water to the spent fuel pool during refueling [5].

For long-term storage, used fuel is sealed inside leak-tight multipurpose canisters and placed in below-ground storage cavities (HI-STORM UMAX). This configuration allows for secure onsite storage while enabling easy retrieval for permanent disposal or reprocessing at a later date [3].

\paragraph{Reactivity Control}
Reactivity in the SMR-300 is regulated through a combination of control rods and soluble boron in the reactor coolant. The control rod assemblies consist of neutron-absorbing materials, such as boron carbide or aluminum carbide, encased within stainless steel or Inconel tubes, in the case of SMR-160's design [6]. These rods are designed to manage reactivity changes during normal operation and ensure safe shutdown when needed.

\paragraph{Safety Features}
While detailed safety system documentation for SMR-300 is not publicly available, it is expected to incorporate key features from the SMR-160 design.

In SMR-160's design, its Reactor Coolant System relies on Primary and Secondary Decay Heat Removal systems to manage decay heat during non-LOCA events. The Primary Decay Heat Removal system consists of two loops: a primary loop, which is part of the reactor coolant pressure boundary and transfers heat from the Reactor Coolant System through a heat exchanger, and a secondary loop, which circulates demineralized water through a second heat exchanger submerged in the Annular Reservoir, where heat is dissipated. The Secondary Decay Heat Removal system provides an additional cooling path by circulating condensate from the steam generator through an external heat exchanger, which is also submerged in the Annular Reservoir, and returning it to the steam generator through the main feedwater system [7].

For normal coolant makeup and small leaks, the Chemical and Volume Control System maintains the reactor coolant inventory by supplying makeup water through two parallel charging pumps. Each pump is designed to compensate for leaks up to the equivalent of a 3/8-inch pipe break in the reactor coolant pressure boundary, which is considered the threshold for LOCA. For leaks up to this threshold, the Chemical and Volume Control System alone is sufficient for plant shutdown without requiring activation of the Passive Core Cooling System. If the charging pumps are unavailable, the Reactor Coolant System will gradually drain, eventually triggering the Passive Core Cooling System due to low pressurizer level or high containment pressure [8]. 

The Passive Core Cooling System of SMR-160 consists of the Passive Core Makeup Water System, which provides safety-related water injection through two medium-pressure accumulators and two low-pressure Passive Core Makeup Water Tanks. To enable coolant injection, the Automatic Depressurization System rapidly reduces Reactor Coolant System pressure, ensuring sufficient flow into the reactor core [8]. The Passive Core Makeup Water Tanks are also connected to the spent fuel pool as a safety-related backup makeup water source, capable of providing up to seven days of cooling in the event of a loss of spent fuel pool cooling. However, if the Passive Core Makeup Water Tanks are needed for core cooling, they will not be used directly for spent fuel pool makeup [9].

\paragraph{Development Timeline}
Holtec International’s SMR-300 (SMR-160) development has progressed through multiple regulatory and project milestones.

In 2020, Holtec was selected for the DOE’s Advanced Reactor Demonstration Program, receiving $\$$147.5 million ($\$$116 million from the DOE) in funding over seven years to advance the SMR-300 design [10,11]. The company has also secured up to $\$$1.52 billion in DOE loan commitments to support efforts in restarting the Palisades Nuclear Generating Station in Michigan, which ceased operations in 2022 [11]. Following the planned restart in 2025, Holtec plans to submit a Construction Permit Application in 2026 for two SMR-300 units at the site, with the first unit targeted for commissioning by mid-2030 [12]. Additionally, Holtec is considering deploying the SMR-300 at its Oyster Creek site in New Jersey to support clean hydrogen production [11,12].

Internationally, Holtec has engaged in the UK’s Generic Design Assessment process, completing Step 1 on August 1, 2024. The reactor has since advanced to Step 2, which is expected to conclude by October 2025 [13].

\paragraph{Reference}
\begin{enumerate}
    \item \textbf{U.S. Nuclear Regulatory Commission}. SMR, LLC (A Holtec International Company). 2025. 
    \item \textbf{Holtech International}. SMR-300 Design Overview. 2024.
    \item \textbf{Holtech International}. Holtec’s SMR-300: A 300-MWe Walk-Away-Safe Nuclear Reactor. 2024.
    \item \textbf{Holtech International}. SMR-300 Core Design Update and Nuclear Analysis Codes and Methods Validation \& Verification Plan. 2024.
    \item \textbf{Holtech International}. NRC Meeting: SFP Makeup. 2022.
    \item \textbf{Holtech International}. Revision 1, SMR 160 Control Rod Assembly Design Criteria Document. 2022.
    \item \textbf{Holtech International}. SMR-160 PDHR and SOHR Containment Isolation White Paper. 2022.
    \item \textbf{Holtech International}. SMR-160 Reactor Coolant Makeup White Paper. 2022.
    \item \textbf{Holtech International}. SMR-160 Spent Fuel Pool Makeup White Paper. 2022.
    \item \textbf{U.S. Department of Energy}. Energy Department’s Advanced Reactor Demonstration Program Awards $\$$30 Million in Initial Funding for Risk Reduction Projects. 2020.
    \item \textbf{U.S. Department of Energy}. Holtec’s Small Modular Reactor Can Go Almost Anywhere, Even Michigan. 2024.
    \item \textbf{Holtec International}. First Two SMR-300 Units Slated to be Built at Michigan’s Palisades Site for Commissioning by Mid-2030. 2023.
    \item \textbf{UK Environment Agency and Natural Resources Wales}. GDA Step 1 of the Holtec SMR: statement of findings public summary. 2024.
\end{enumerate}

\clearpage
\newpage
\subsection{BWRX-300 (GE-Hitachi)}

\begin{wraptable}{o}{9cm} 
    \vspace{-12mm} 
    \centering
    \begin{threeparttable} 
        \renewcommand{\arraystretch}{1.2} 
        \rowcolors{2}{gray!15}{white} 
        \begin{tabular}{ p{4cm} | p{4cm} } 
            \toprule
            \rowcolor{white}
            \multicolumn{2}{l}{\textbf{General Information}} \\ 
            \midrule
            Reactor Type & Boiling Water Reactor \\
            Purpose & Commercial \\
            Thermal Power (MWt) & 870 \\
            Net Power Output (MWe) & 300 \\
            Design Life & 60 years \\
            Reactor Units per Site & Multiple (4) units \\
            Seismic Design & 0.3 g \\ 
            Site Footprint (m\textsuperscript{2}) & 27,000 \\
            Construction Time (NOAK) & -\\
            
            \midrule
            \rowcolor{white}
            \multicolumn{2}{l}{\textbf{Fuel $\&$ Materials}} \\ 
            \midrule
            
            Core Coolant & H\textsubscript{2}O \\
            Neutron Moderator & H\textsubscript{2}O \\
            Solid Burnable Absorber  & B\textsubscript{4}C, Hf, Gd\textsubscript{2}O\textsubscript{3} \\
            Fuel Cladding & Zircaloy-2 \\
            Fuel Material & UO\textsubscript{2} \\
            Fuel Enrichment & avg 3.8\%, max 4.95\% \\
            Refueling Cycle & 12 to 24 months\\
            
            \midrule
            \rowcolor{white}
            \multicolumn{2}{l}{\textbf{Development $\&$ Licensing}} \\ 
            \midrule
            
            Design Status & Detailed Design\\
            Licensing Status & Preliminary Safety Analysis Report \\
            \bottomrule
        \end{tabular}

        \begin{tablenotes}
            \item {Last ARIS update on 2024/10/11}
        \end{tablenotes}
    \end{threeparttable}
    \vspace{-10mm} 
\end{wraptable}

Although numerous topical reports and white papers have been submitted to the NRC, most design information on the BWRX-300 is sourced from the BWRX-300 General Description document (Revision F) by GE Hitachi Nuclear Energy, as it is the most recent publication (December 2023).

\begin{figure}[b]
    \centering
    \includegraphics[width=0.8\linewidth]{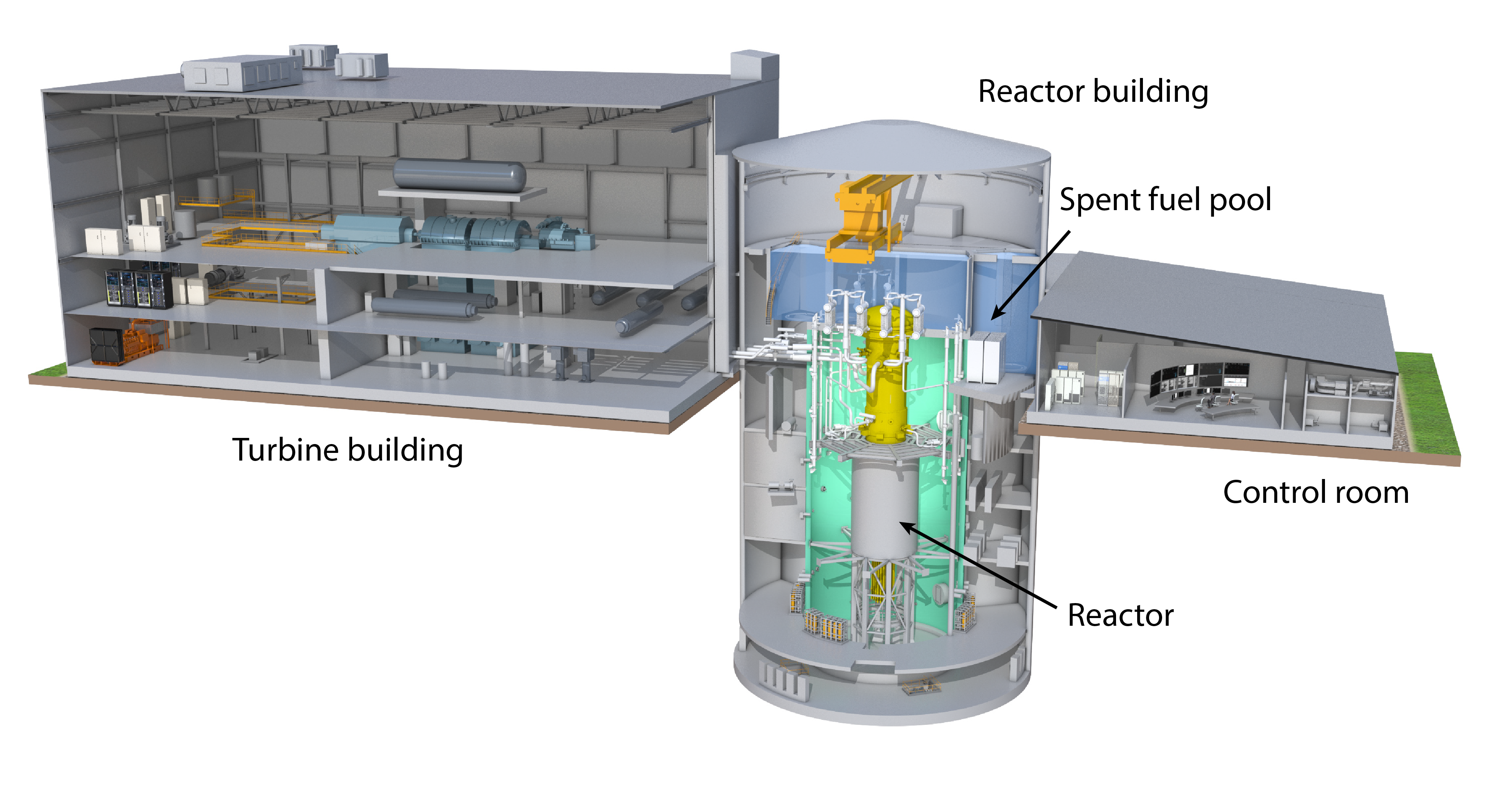}
    \caption{Plant design of the BWRX-300 SMR. Image Credit: Hitachi-GE Nuclear Energy.}
    \label{fig:bwrx-300}
\end{figure}

\paragraph{Basic Design}
The BWRX-300 is a 300 MWe SMR developed by GE-Hitachi Nuclear Energy, designed for flexible deployment in electricity generation, industrial applications, and hydrogen production. Non-electric applications include process heat supply, with an available temperature range of 100–260\textsuperscript{o}C [1]. It is a scaled-down version of GE-Hitachi's Economic Simplified Boiling Water Reactor and incorporates passive safety features to enhance reliability and leverages commercial off-the-shelf equipment to achieve cost-effectiveness. Figure \ref{fig:bwrx-300} provides an overview of the plant design of the BWRX-300.

BWRX-300's Reactor Pressure Vessel features a tall chimney, enabling natural circulation cooling, eliminating the need for recirculation pumps while ensuring stable core cooling [2]. Planned refueling cycles range from 12 to 24 months, with major equipment inspections every 120 months [3].

\paragraph{Fuel Design}
The BWRX-300 core consists of 240 fuel assemblies, using Global Nuclear Fuel’s GNF2 fuel design. Each 10$\times$10 fuel assembly contains 92 fuel rods (78 full-length, 14 part-length) and two central water rods [4]. The fuel is low-enriched UO\textsubscript{2}, with an average enrichment of 3.81\% and a maximum of 4.95\% U-235 [1]. 

The reactor’s fuel handling system allows for short-term and long-term spent fuel storage in the reactor building, with the storage racks having a total capacity of approximately 275\% of a complete core load. Additionally, used fuel can be transferred to dry cask storage in an on-site Independent Spent Fuel Storage Installation facility for long-term storage [3].

\paragraph{Reactivity Control}
The BWRX-300 employs Fine Motion Control Rod Drives for precise reactivity management during normal operation. The 57 control rods are distributed throughout the reactor core and contain neutron-absorbing materials, including boron carbide powder and hafnium rods [5]. In emergency conditions, the cruciform-shaped control rods are rapidly insert into the core, ensuring a swift reactor shutdown. In addition, burnable absorbers, such as gadolinium oxide, are incorporated into the fuel pellets to help regulate excess reactivity throughout the fuel cycle [5]. 

As a backup shutdown mechanism, the Boron Injection System ensures reactor sub-criticality from full-power operation in the event of control rod system failure, providing an additional layer of safety [6].

\paragraph{Safety Features}
The BWRX-300 employs multiple passive safety systems to enhance resilience against design-basis accidents and external events. The Primary Containment System consists of a steel-plate composite containment vessel, measuring 17.5 meters in diameter and 38 meters in height [7]. This containment structure is designed to maintain structural integrity during severe accident scenarios, preventing the release of radioactive materials. The containment atmosphere is nitrogen-inerted to dilute hydrogen and oxygen concentrations, minimizing the risk of hydrogen explosion following an accident [7].

The Containment Cooling System and Passive Containment Cooling System regulate containment temperature and pressure. Under normal conditions, Containment Cooling System actively cools the containment, while the Passive Containment Cooling System contributes little to heat removal. However, in the event of a pipe breach where steam is discharged into the containment, the Passive Containment Cooling System manages containment temperature and pressure using natural circulation and condensation mechanisms. It consists of three independent low-pressure heat exchangers, which transfer heat from the containment to an above-containment equipment pool, which is filled with water [8]. 

During accidents, the pressure vessel is cooled using the Isolation Condenser System, which consists of three independent heat exchanger trains, each submerged in a dedicated water pool. With two Isolation Condenser System trains in operation, the system can provide stable pressure vessel cooling for at least seven days without operator action and can function indefinitely if the water pool is replenished [9].

Finally, the Fuel Pool Cooling and Cleanup System ensures continuous cooling of spent fuel through two independent sets of pumps, demineralizers, and heat exchangers. A single set is sufficient to prevent bulk boiling in the fuel pool. In the event of a failure of both sets, the pool is designed to maintain sufficient water coverage for at least seven days [10].

\paragraph{Development Timeline}
GE-Hitachi Nuclear Energy has pursued licensing for the BWRX-300 across multiple countries. In Canada, the Vendor Design Review Phases 1 and 2 were completed in March 2023, with the Canadian Nuclear Safety Commission confirming no fundamental barriers to licensing [11]. A License to Construct application was submitted by Ontario Power Generation to the Canadian Nuclear Safety Commission in October 2022 for the Darlington site, with the results of the latest regulatory hearing planned for January 2025 not yet available as of March 2025 [12]. 

In the United Kingdom, GE-Hitachi entered the Generic Design Assessment process in December 2022, advancing to Step 2 by late 2024 [13]. Step 2 is expected to be finished by December 2025, but no Design Acceptance Confirmation or Statement of Design Acceptability will be issued at the end of the currently agreed program of work [13].

In the United States, GE-Hitachi has engaged in pre-application activities with the NRC since 2019, submitting multiple Licensing Topical Reports and White Papers. While GE-Hitachi anticipated submitting a Construction Permit application in 2024, no public confirmation of this has been made [14]. 

Beyond North America and the UK, GE-Hitachi and its partners are in pre-application discussions with Poland’s National Atomic Energy Agency and the Czech Republic’s State Office for Nuclear Safety and Radioactive Waste Repository Authority, exploring the use of Canadian or U.S. regulatory approvals to streamline licensing in these countries [14]. BWRX-300 was also selected by Fermi Energia for future deployment in Estonia by early 2030s [15].

\paragraph{Reference}
\begin{enumerate}
    \item \textbf{GE Hitachi Nuclear Energy}. Appendix A - BWRX-300 Parameters of Interest, BWRX-300 General Description. 2023.
    \item \textbf{International Atomic Energy Agency}. Small Modular Reactor Technology Catalogue 2024 Edition, Second edition. 2025. 
    \item \textbf{GE Hitachi Nuclear Energy}. Section 4.3 Fuel Handling and Refueling Process, BWRX-300 General Description. 2023.
    \item \textbf{GE Hitachi Nuclear Energy}. Section 4.2 Fuel Assembly Description, BWRX-300 General Description. 2023.
    \item \textbf{GE Hitachi Nuclear Energy}. BWRX-300 Reactivity Control. 2020.
    \item \textbf{GE Hitachi Nuclear Energy}. Section 3.7 Boron Injection System, BWRX-300 General Description. 2023.
    \item \textbf{GE Hitachi Nuclear Energy}. Section 3.4 Containment, BWRX-300 General Description. 2023.
    \item \textbf{GE Hitachi Nuclear Energy}. Section 3.6 Passive Containment Cooling System, BWRX-300 General Description. 2023.
    \item \textbf{GE Hitachi Nuclear Energy}. Section 3.3 Isolation Condenser System, BWRX-300 General Description. 2023.
    \item \textbf{GE Hitachi Nuclear Energy}. Section 3.11 Fuel Pool Cooling and Cleanup System, BWRX-300 General Description. 2023.
    \item \textbf{World Nuclear News}. BWRX-300 completes Phases 1 \& 2 of Canadian pre-licensing review. 2023.
    \item \textbf{Canadian Nuclear Safety Commission}. New reactor facility projects. 2024.
    \item \textbf{UK Environment Agency, Office for Nuclear Regulation and Natural Resources Wales}. GE-Hitachi’s Small Modular Reactor completes first step of design assessment. 2024.
    \item \textbf{GE Hitachi Nuclear Energy}. Section 1.6 BWRX-300 Status, BWRX-300 General Description. 2023.
    \item \textbf{World Nuclear News}. BWRX-300 selected for Estonia's first nuclear power plant. 2023.
\end{enumerate}

\clearpage
\newpage
\subsection{Xe-100 (X-energy)}

\begin{wraptable}{o}{8cm} 
    \vspace{-12mm} 
    \centering
    \begin{threeparttable} 
        \renewcommand{\arraystretch}{1.2} 
        \rowcolors{2}{gray!15}{white} 
        \begin{tabular}{ p{4cm} | p{3cm} } 
            \toprule
            \rowcolor{white}
            \multicolumn{2}{l}{\textbf{General Information}} \\ 
            \midrule
            Reactor Type & Gas-Cooled Reactor \\
            Purpose & Commercial \\
            Thermal Power (MWt) & 200 \\
            Net Power Output (MWe) & 82.5 \\
            Design Life & 60 years \\
            Reactor Units per Site & - \\
            Seismic Design & 0.5 g \\ 
            Site Footprint (m\textsuperscript{2}) & - \\
            Construction Time (NOAK) & - \\
            
            \midrule
            \rowcolor{white}
            \multicolumn{2}{l}{\textbf{Fuel $\&$ Materials}} \\ 
            \midrule
            
            Core Coolant & He \\
            Neutron Moderator & Graphite \\
            Solid Burnable Absorber  & - \\
            Fuel Cladding & - \\
            Fuel Material & UCO \\
            Fuel Enrichment & 15.5\% \\
            Refueling Cycle & Continuous \\
            
            \midrule
            \rowcolor{white}
            \multicolumn{2}{l}{\textbf{Development $\&$ Licensing}} \\ 
            \midrule
            
            Design Status  & Basic Design \\
            Licensing Status & -   
            \end{tabular}

        \begin{tablenotes}
            \item {Last ARIS update on 2024/10/17}
        \end{tablenotes}
    \end{threeparttable}
    \vspace{-30mm} 
\end{wraptable}

\paragraph{Basic Design}
The Xe-100 is a Generation IV high-temperature gas-cooled reactor that utilizes a pebble-bed core and helium coolant to generate 200 MWt per unit. Each unit produces 80 MWe with a thermal efficiency of 40\%, and a standard plant consists of four independently-operating units, providing a total output of 320 MWe [1].

The reactor employs TRISO-coated uranium oxycarbide (UCO) fuel in spherical pebbles, enhancing fuel integrity and safety. UCO refers to the mixture of uranium dioxide and uranium carbide phases present in the fuel kernel. Heat is transferred via helium to a helical-coil steam generator, producing high-quality superheated steam at 565°C and 16.5 MPa, enabling applications beyond electricity generation, including industrial process heat, district heating, and cogeneration [1,2].

\begin{figure}[b!]
    \centering
    \includegraphics[width=0.6\linewidth]{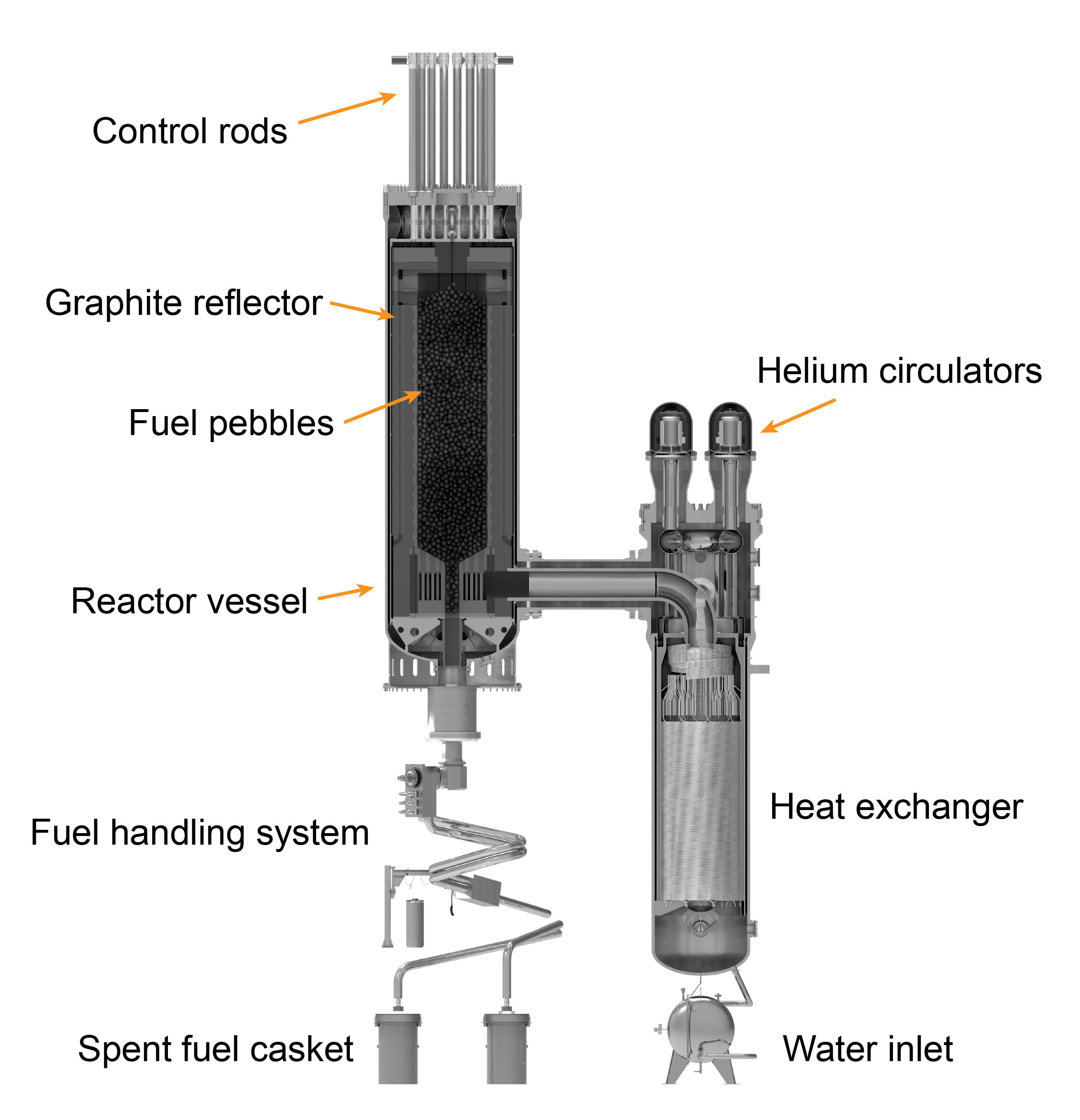}
    \caption{Section view of the Xe-100 reactor. The components are labeled based on a licensing topical report submitted by X-energy. The original image was provided by X-energy.}
    \label{fig:xe-100-section-view}
\end{figure}

The Xe-100's plant design consists of a Nuclear Island housing the reactor and its paired steam generator, together with all safety-related systems, and a Conventional Island that contains commercially available steam turbine systems [1,2]. A section view of the Xe-100’s reactor and steam generator is presented in Figure \ref{fig:xe-100-section-view}.

\newpage
\paragraph{Fuel Design}
The Xe-100’s fuel design revolves around its pebble-bed core, which accommodates approximately 224,000 spherical fuel elements known as pebbles [3]. This design enables online refueling, allowing the reactor to operate with minimal excess reactivity and achieve high availability levels above 95\% [1]. During refueling, fresh pebbles are introduced into the reactor through a central tube at the top of the pressure vessel and flow downward through the core. Simultaneously, used fuel pebbles exit via a discharge system and undergo burnup measurement using gamma spectroscopy. Based on their burnup level, they are either sent for long-term storage or reinserted into the reactor for additional cycles, typically completing around six cycles before reaching their burnup limit [3]. Approximately 175 new fuel pebbles are loaded into and removed from the reactor per day [1]. The spent fuel pebbles are stable enough to be stored in spent fuel casket, eliminating the need for an actively cooled spent fuel pool [2]. The fuel handling system and spent fuel casket are labeled in Figure \ref{fig:xe-100-section-view}.

\begin{figure}[h]
    \centering
    \includegraphics[width=0.8\linewidth]{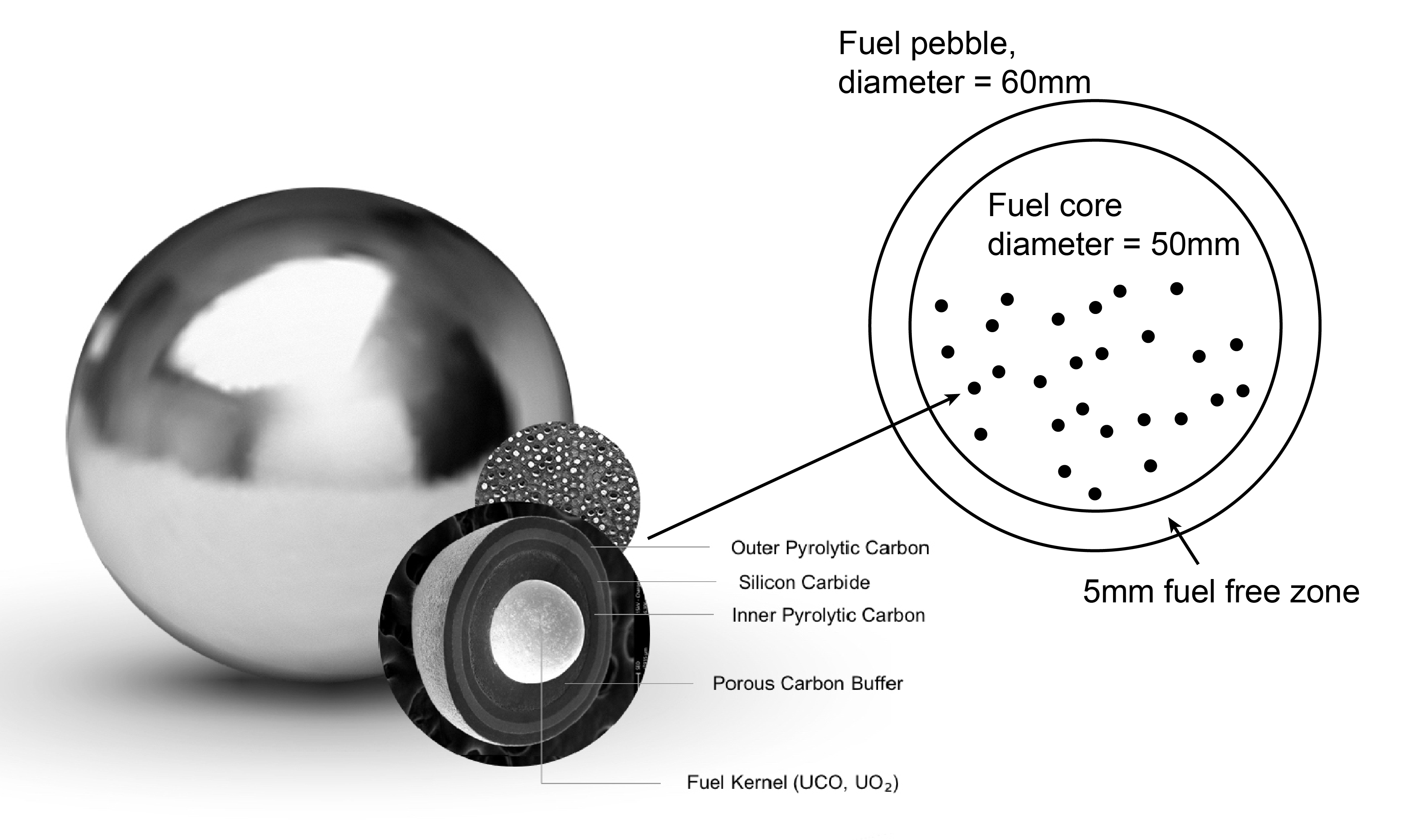}
    \caption{Image of the TRISO fuel pebble. The dimensions of the fuel pebbel, the fuel free zone, and the fuel core are labeled based on a licensing topical report submitted by X-energy. The original image was provided by X-energy.}
    \label{fig:triso-pebble}
\end{figure}

As shown in Figure \ref{fig:triso-pebble}, each fuel pebble measures 60 mm in diameter and consists of a 50 mm fuel core surrounded by a 5 mm fuel-free zone [3]. Each fuel pebble contains approximately 19,000 TRISO-coated particles embedded in a carbonaceous matrix. The TRISO particles are comprised of UCO fuel kernels enriched to 15.5\%, with a diameter of 425 µm [3]. Each kernel is surrounded by multiple layers of pyrolytic carbon and a silicon carbide (SiC) coating, forming a miniature multi-shell pressure vessel. The TRISO coatings ensure high fuel integrity by retaining a substantial fraction of fission products within the kernel, acting as a primary barrier against radionuclide release.

The fuel pebbles are fully wetted by the helium coolant, which leads to low temperature gradients across the fuel particles and lower maximum fuel temperatures during normal operation [3]. The maximum fuel temperature is limited to well below 1000°C under normal operating conditions [4].

\paragraph{Reactivity Control}
The Xe-100’s reactivity control is primarily achieved through its inherent design features and the use of two independent control rod systems. The core's excess reactivity is limited by the capability for online refueling, allowing fuel to be loaded and unloaded as needed during full power operation. Additionally, the reactor has an overall negative temperature coefficient of reactivity driven by Doppler broadening of the fuel kernel content and the graphite moderator’s temperature coefficient [3]. This ensures the reactor will inherently shut itself down to a hot standby condition during off-normal events when forced cooling is lost [5]. 

While the negative temperature coefficient provides the primary reactivity control, there is a slight delay due to the need for heat-up to occur before reactivity is reduced. To mitigate this delay and maintain stable steam temperatures for the steam turbine, the Reactivity Control System is used, which consists of nine control rods made from Incoloy 800H and boron carbide [3,5]. These rods are inserted into cylindrical channels in the side reflector close to the pebble-bed core to adjust reactivity during normal operation and achieve a hot standby if required.

In addition to the Reactivity Control System, the reactor is equipped with a Reserve Shutdown System, which provides a second, diverse means of negative reactivity insertion. The shutdown system is also composed of nine control rods that are identical in design to those of the control system [3]. The Reserve Shutdown System is operated by the safety-related Reactor Protection System and is used to achieve and maintain a cold shutdown condition indefinitely. Although each system is capable of independently shutting down the reactor to a hot standby condition, both are required to bring the reactor into a cold shutdown state [4,5].

\paragraph{Safety Features}
The Xe-100 is designed to ensure inherent safety through its core components and passive safety mechanisms that do not require external power or operator intervention to stabilize fuel temperatures. It employs a unique safety approach that does not rely on a conventional leak-tight containment structure. Instead, it utilizes multiple layers of defense to retain radionuclides within the reactor and meet regulatory dose limits. The containment approach consists of five principal barriers: (1) the fuel kernel, (2) the TRISO particle coatings, particularly the SiC layer, (3) the spherical fuel-element matrix, including the fuel-free zone, (4) the reactor helium pressure boundary, and (5) the reactor building. Each of these barriers plays a critical role in limiting radionuclide release, even in beyond-design-basis events [2,5].

The first three barriers are provided by the fuel pebble structure, including its fuel-free zone. UCO TRISO-coated fuel offers over 99.99\% fission product retention during both normal operation and beyond-design-basis events, with demonstrated retention capability at temperatures up to 1800°C for 300 hours [5]. Both the UCO fuel kernel and the SiC layer are essential in containing fission products. The buffer layer, composed of low-density pyrolytic carbon, accommodates gaseous fission products, reducing internal pressure buildup and allowing for fuel kernel swelling under irradiation. The inner and outer pyrolytic carbon layers shrink under irradiation, creating compressive stress that counteracts the tensile forces generated by fission gas buildup, ensuring the structural integrity of the fuel particle [4,5].

Any radionuclides that escape from the fuel can circulate within the primary helium coolant system, which includes monitoring and active filtration capabilities. This system continuously measures and removes contamination, keeping radionuclide levels within operational limits during normal operation and minimizing releases after an accident [5].

The reactor building serves as the final passive barrier against fission product release but is not the primary structure for containing radionuclides, as most remain trapped within the fuel. To maintain its structural integrity, the concrete reactor building is kept at controlled temperatures by the Reactor Cavity Cooling System. Additionally, the building features an overpressure relief system, which passively vents to the atmosphere in the event of a large helium or steam piping failure, preventing excessive internal pressure buildup [5]. 

The Reactor Cavity Cooling System consists of two trains of tube curtains located outside the pressure vessel designed to operate in two modes: an active cooling mode during normal operation and a passive boil-off mode in accident conditions [5]. When active cooling is lost during a design basis event or design basis accident, the system transitions to its safety-related passive mode, which provides at least 72 hours of heat removal through water boil-off [5]. If additional cooling is required beyond this period, onsite operators can replenish the water supply to sustain heat removal. In addition to water-based cooling, the Reactor Cavity Cooling System also provides a passive heat transfer pathway, directing decay heat away from the reactor vessel through a combination of conduction, natural convection, and radiation into the ground through the concrete reactor building [4,5].

Together with the Reactor Cavity Cooling System, numerous additional design features of the Xe-100 contribute to its inherent safety [2,5]. Notably, the reactor has a low core power density, which helps limit temperature rise during off-normal conditions, and its excess reactivity can be continuously adjusted through online refueling. Furthermore, the large negative fuel and moderator temperature coefficients inherently limit power increases and gradually reduce reactor power under accident conditions without requiring control rods insertion.

The reactor’s large solid graphite moderator and reflector structure enhances inherent safety by providing substantial thermal inertia, delaying core thermal transients for hours or even days compared to conventional light water reactors. Moreover, the single-phase, chemically inert helium coolant has a low heat capacity, meaning it retains little thermal energy. This minimizes safety concerns in the event of a helium pressure boundary breach.

\paragraph{Development Timeline}
In 2015, X-energy received the Advanced Reactor Concepts award from the DOE to further develop the Xe-100 reactor design and TRISO-X fuel [6]. Since September 2018, the company has engaged in pre-application activities with the NRC, submitting multiple topical reports and white papers. Progress accelerated in 2020 when X-energy was selected for the DOE Advanced Reactor Demonstration Program, securing an initial $\$$80 million, with a total of $\$$1.23 billion over seven years, to deploy a four-unit, 320 MWe Xe-100 plant by 2027 [7]. The funding also supports the TRISO-X fuel fabrication facility, which broke ground in Oak Ridge, Tennessee, in October 2022 [8].

In 2023, X-energy signed a Joint Development Agreement with Energy Northwest to deploy up to 12 Xe-100 modules (960 MWe) in Washington, with the first unit expected online by 2030 [9]. Separately, X-energy is applying new NRC guidance to its Construction Permit application for an Xe-100 plant in Seadrift, Texas, with submission planned for 2024, though no confirmation has been made as of March 2025 [10].

Internationally, X-energy completed Phases 1 and 2 of the Canadian Nuclear Safety Commission's Vendor Design Review by January 2024, with the commission finding no fundamental barriers to licensing [11]. The company has indicated plans to pursue Phase 3 Vendor Design Review [11]. In April 2024, X-energy and Cavendish Nuclear received GBP 3.34 million (USD 4.23 million) in UK government funding to support the Xe-100’s deployment, including a 12-reactor plant at Hartlepool, targeting completion in the early 2030s [12].

\paragraph{Reference}
\begin{enumerate}
    \item \textbf{X-energy}. Xe-100 White Paper: Physical Protection System Approach. 2022.
    \item \textbf{Southern Company}. X-energy Xe-100 TICAP Tabletop Exercise Report. 2021. 
    \item \textbf{X-energy}. Xe-100 Licensing Topical Report: Reactor Core Design Methods and Analysis. 2024.
    \item \textbf{X-energy}. Xe-100 Topical Report: TRISO-X Pebble Fuel Qualification Methodology. 2021.
    \item \textbf{X-energy}. Xe-100 Licensing Topical Report Transient and Safety Analysis Methodology. 2024.
    \item \textbf{U.S. Department of Energy}. X-energy Completes $\$$40 Million Project to Further Develop High-Temperature Gas Reactor. 2022.
    \item \textbf{X-energy}. X-energy signs Department of Energy’s Advanced Reactor Demonstration Program (ARDP) Cooperative Agreement. 2021.
    \item \textbf{X-energy}. TRISO-X Breaks Ground on North America’s First Commercial Advanced Nuclear Fuel Facility. 2022.
    \item \textbf{X-energy}. Energy Northwest and X-energy Sign Joint Development Agreement for Xe-100 Advanced Small Modular Reactor Project. 2023. 
    \item \textbf{U.S. Department of Energy}. NRC Endorses New Guidance for Advanced Reactor Licensing. 2024. 
    \item \textbf{X-energy}. X-energy Successfully Completes Canadian Pre-Licensing Milestone for the Xe-100 Advanced Small Modular Reactor. 2024.
    \item \textbf{X-energy}. UK Government Selects X-energy and Cavendish Nuclear for First Advanced Modular Reactor Award from Future Nuclear Enabling Fund. 2024. 
\end{enumerate}

\clearpage
\newpage
\subsection{KP-FHR (Kairos Power)}

\begin{wraptable}{o}{9cm} 
    \vspace{-12mm} 
    \centering
    \begin{threeparttable} 
        \renewcommand{\arraystretch}{1.2} 
        \rowcolors{2}{gray!15}{white} 
        \begin{tabular}{ p{4cm} | p{4cm} } 
            \toprule
            \rowcolor{white}
            \multicolumn{2}{l}{\textbf{General Information}} \\ 
            \midrule
            Reactor Type & Molten Salt Reactor \\
            Purpose & Demonstration \\
            Thermal Power (MWt) & 320 \\
            Net Power Output (MWe) & 140 \\
            Design Life & 80 years \\
            Reactor Units per Site & Single unit \\
            Seismic Design & - \\ 
            Site Footprint (m\textsuperscript{2}) & - \\
            Construction Time (NOAK) & - \\
            
            \midrule
            \rowcolor{white}
            \multicolumn{2}{l}{\textbf{Fuel $\&$ Materials}} \\ 
            \midrule
            
            Core Coolant & Molten salts \\
            Neutron Moderator & Graphite \\
            Solid Burnable Absorber  & - \\
            Fuel Cladding & - \\
            Fuel Material & TRISO particles in graphite matrix \\
            Fuel Enrichment & 19.75\% \\
            Refueling Cycle & Continuous \\
            
            \midrule
            \rowcolor{white}
            \multicolumn{2}{l}{\textbf{Development $\&$ Licensing}} \\ 
            \midrule
            
            Design Status & Conceptual Design \\
            Licensing Status & -     
            \end{tabular}

        \begin{tablenotes}
            \item {Last ARIS update on 2024/10/15}
        \end{tablenotes}
    \end{threeparttable}
    \vspace{-5mm} 
\end{wraptable}

\paragraph{Basic Design}
The Kairos Power - Fluoride High-Temperature Reactor (KP-FHR) is a Generation IV high-temperature reactor that utilizes molten fluoride salt as its coolant. Kairos Power has developed the KP-FHR by building upon more than a decade of research conducted at U.S. national laboratories and universities on FHR designs [1]. According to Kairos Power, a commercial-scale KP-FHR plant is designed to feature two 75 MWe reactor units, producing a total net electrical output of 150 MWe [2].

The KP-FHR operates at near-atmospheric pressure and utilizes TRISO-coated fuel particles, originally designed for high-temperature gas-cooled reactors. These fuel particles are embedded within a pebble fuel element, serving as the primary containment barrier for fission products [1,3].

\begin{figure}[h]
    \centering
    \includegraphics[width=0.9\linewidth]{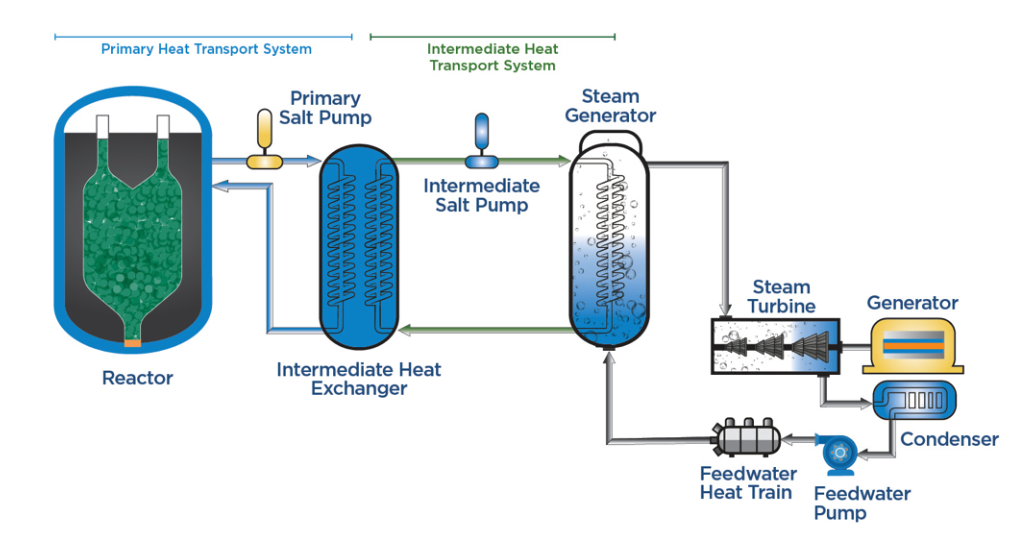}
    \caption{Schematic of the Kairos Power KP-FHR power generation system. Image Credit: Kairos Power.}
    \label{fig:kp-fhr-plant}
\end{figure}

As illustrated in Figure \ref{fig:kp-fhr-plant}, the KP-FHR employs a two-loop heat transport system, where the primary coolant loop transfers heat to an intermediate coolant loop that utilizes a compatible nitrate salt (mixture of NaNO\textsubscript{3}-KNO\textsubscript{3}) [3]. This intermediate loop then delivers heat to the power conversion system via a steam generator. The reactor coolant is a chemically stable molten fluoride salt mixture referred to as Flibe (2LiF:BeF\textsubscript{2}), which not only facilitates heat transfer but also provides an additional fission product retention barrier [1].

The KP-FHR features an online refueling design, which allows for continuous adjustments to control the core's reactivity [2]. It also incorporates passive decay heat removal, eliminating the need for electrical power to maintain safe shutdown conditions in the event of an accident [1].

\begin{figure}[h]
    \centering
    \includegraphics[width=0.85\linewidth]{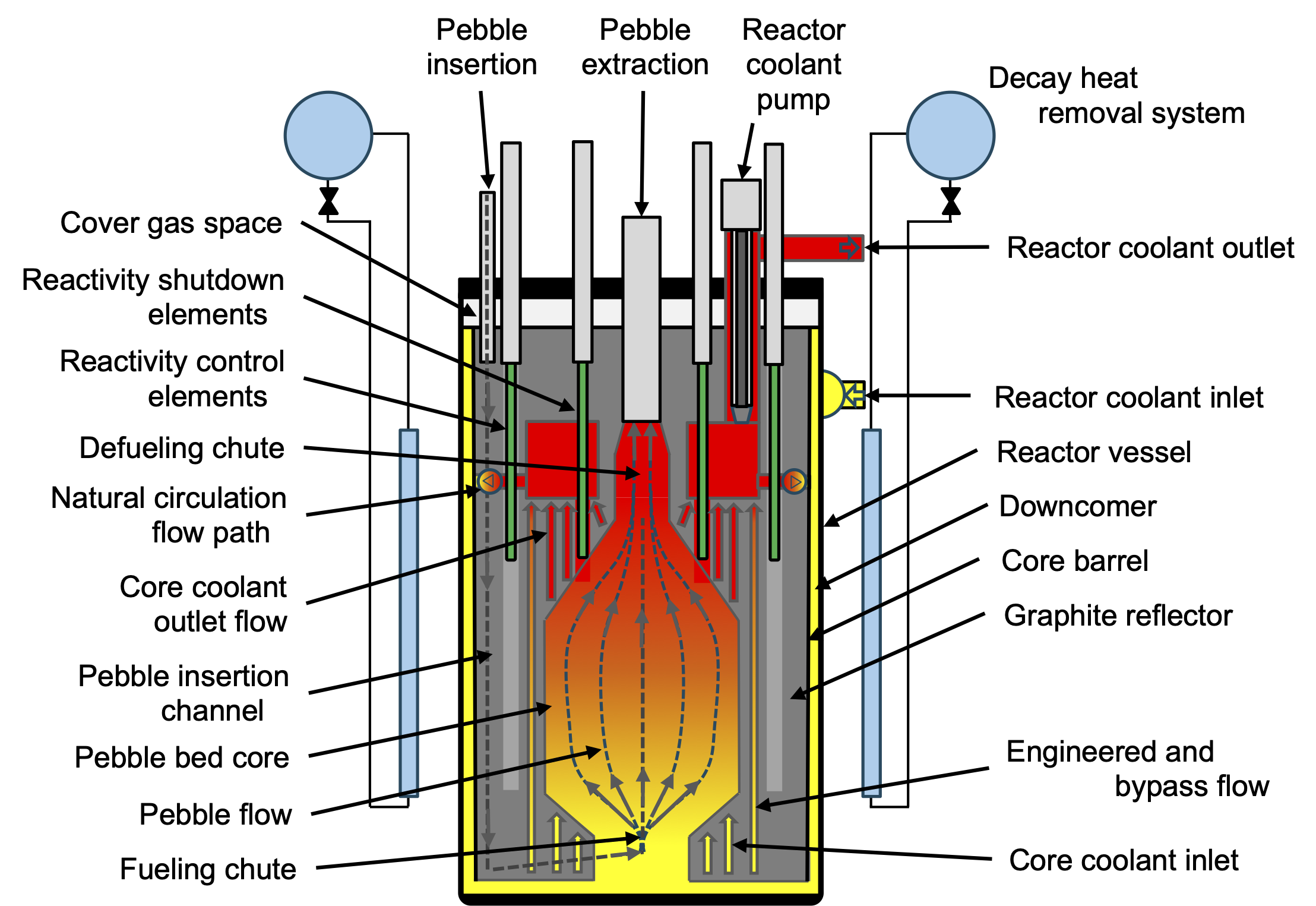}
    \caption{Detailed illustration of the KP-FHR reactor components. Image Credit: Kairos Power}
    \label{fig:kp-fhr-core}
\end{figure}

\paragraph{Fuel Design}
The KP-FHR reactor vessel is constructed from 316H stainless steel and consists of a cylindrical outer wall, core barrel, and disc-shaped upper head, as shown in Figure \ref{fig:kp-fhr-core}. The vessel provides the primary flow path for reactor coolant during normal operations, where the coolant enters through the side inlet, flows downward through the downcomer annulus, and is guided through the core by the graphite reflector blocks [4,5].

The KP-FHR core is a randomly packed pebble-bed, surrounded by stacked graphite reflector blocks. These blocks serve as neutron reflectors and contain engineered channels for directing coolant, housing instrumentation, and inserting control elements. The graphite blocks, along with the fuel pebbles, are buoyant in the molten fluoride salt coolant, which operates at high temperature and near-atmospheric pressure [1].

The fuel pebbles in KP-FHR use the UCO TRISO-coated fuel design, embedded in a spherical graphite matrix. The components of the TRISO-coated fuel particles are described in the Xe-100 section. However, unlike Xe-100, where TRISO fuel particles are distributed throughout the central fuel core, KP-FHR fuel pebbles have a low-density inner core to maintain buoyancy in the coolant, with TRISO particles concentrated in an annular shell. An additional fuel-free carbon matrix shell surrounds the fuel annulus to protect it from mechanical damage [1,5,6].

The fuel particles can have a range of enrichments, from depleted uranium up to the upper limit of high-assay low-enriched uranium (20 wt\% U-235) [1].  The core may also contain graphite-only moderator pebbles that have the same diameter as the fuel pebbles, contain no uranium, and are made entirely of the same graphite matrix material used in the fuel pebbles, but lack an inner low-density core. The ability to adjust the mixture of pebble types allows for reactivity control during startup and core transitions [1,5]. In addition to moderator pebbles, neutron moderation in KP-FHR is provided by the graphite in the fuel pebbles, the Flibe coolant, and the graphite reflector blocks [5].

Refueling operations are managed by the pebble handling and storage system to regulate core reactivity. As shown in Figure \ref{fig:kp-fhr-core}, fresh pebbles enter the core through the pebble insertion line at the bottom of the reactor, while spent pebbles are removed from the top of the core via the pebble extraction machine. Extracted pebbles undergo burnup and damage evaluation, after which they are either reinserted into the core or sent to storage [1,5].

\paragraph{Reactivity Control}
The Reactivity Control and Shutdown System consists of control elements and shutdown elements, both primarily composed of B\textsubscript{4}C neutron absorber material, enclosed within stainless steel 316H cladding. The B\textsubscript{4}C absorber material is formed into pellets and stacked within cylindrical SS 316H tubes pressurized with inert gas. [5]

The Reactivity Control and Shutdown System design includes seven elements in total: four control elements and three shutdown elements, each with distinct designs and insertion locations. The non-safety-related control elements regulate reactivity during normal startup, power adjustments, and planned operational changes, while the safety-related shutdown elements are responsible for reactor shutdown during off-normal conditions.  In the event of a reactor trip, both control and shutdown elements are automatically released, dropping under gravity into the side graphite reflector and directly into the pebble bed, respectively, to ensure the safe shutdown of the plant [1,5].

\paragraph{Safety Features}
The functional containment of the KP-FHR relies on three key physical barriers to limit the release of radioactive material: TRISO-coated fuel, the Flibe coolant, and the reactor vessel [5]. Similar to the Xe-100 design, the TRISO-coated fuel particles serve as the primary containment for KP-FHR, with a design temperature of 1600\textsuperscript{o}C, providing a significant safety margin and multiple protective layers that retain fission products under normal and off-normal conditions. The UCO fuel kernel and the SiC layer are the most critical barriers preventing the release of radioactivity, with other layers also contributing to radionuclide retention while providing structural support [5].

The fuel’s failure modes are classified as either TRISO failure (exposed kernel) or SiC failure. A TRISO failure occurs when all outer layers are compromised, leading to the release of fission gases. In contrast, a SiC failure occurs when the SiC layer is damaged, resulting in the detection of cesium, but at least one dense pyrolytic carbon layer remains intact to help trap fission gas in the fuel particle [5,6].

Beyond the TRISO-coated fuel, the Flibe coolant acts as an additional barrier, retaining fission products that may escape from the fuel. During normal operations, the non-safety-related chemistry control system monitors coolant chemistry through periodic sampling and manages coolant replacement via the Inventory Management System [4,5]. 

Finally, the reactor vessel, which operates at near-atmospheric pressure, effectively prevents the energetic releases typically associated with highly pressurized systems, further enhancing containment integrity. During a postulated event where the heat transport systems are unavailable, the reactor vessel internals provide an alternative natural circulation path, as shown in Figure \ref{fig:kp-fhr-core}, to facilitate heat removal from the reactor core. In such a scenario, the reactor coolant naturally circulates from the core to the downcomer annulus, transferring heat to the reactor vessel wall, which is subsequently cooled by the Decay Heat Removal System [5]. 

The Decay Heat Removal System is a fully passive ex-vessel cooling system designed to operate continuously when the reactor is above a threshold power level under both normal and off-normal conditions. The system requires no external electrical power or operator intervention during postulated events. It consists of four independent cooling trains, each comprising a water storage tank, steam separator, and six thimbles positioned along the outer surface of the reactor vessel to ensure system redundancy and reliability [5]. 

At low reactor power levels, natural parasitic heat loss is sufficient to cool the reactor vessel, and the thimbles remain dry. However, when decay heat exceeds the cooling capacity of passive heat loss, the isolation valves of the thimble feedwater lines open, causing the water storage tanks to flood the thimbles through the steam separator. The water inside the thimbles boils off, transferring heat away from the reactor vessel walls, and the generated steam is passively vented into the atmosphere [5]. 

As the water boils off in the thimble, the steam separator is continuously replenished from the water storage tanks via gravity, sustaining long-term cooling. For the Decay Heat Removal System to function effectively during postulated events, at least three of the four water storage tanks must be operational. Each tank contains a sufficient water inventory to sustain continuous operation of its connected thimbles for up to seven days, even if the primary feedwater supply is lost [5]. 

Finally, two biological shields form barriers to protect plant workers and the public from radiological exposure. The primary biological shield is constructed of concrete and is located just outside the reactor vessel, housing the primary salt pump. The secondary biological shield is located outside the primary biological shield and contains the Inventory Management System and the Intermediate Heat Exchanger shown in Figure \ref{fig:kp-fhr-plant} [5].

\paragraph{Development Timeline}
Kairos Power’s KP-FHR development has progressed through several key milestones, with a primary focus on deploying Hermes low-power demonstration reactors in the United States.

In 2020, Kairos Power was selected for the DOE’s Advanced Reactor Demonstration Program to support the design, construction, and commissioning of the Hermes reduced-scale test reactor [7]. The company submitted a Construction Permit Application for Hermes with the NRC on September 29, 2021, receiving NRC approval on December 14, 2023 [8]. This milestone made Hermes the first Generation IV reactor to receive an NRC construction permit, though it is a non-electric test reactor [8,9]. Kairos Power also submitted the Construction Permit Application for Hermes 2 in July 2023 [10].

2024 saw several major developments. In February, the company signed a Technology Investment Agreement with the DOE, finalizing the Advanced Reactor Demonstration Program award for $\$$303 million under a performance-based, fixed-price milestone approach [11]. In July, Kairos Power began construction of the Hermes reactor in Oak Ridge, Tennessee, with operations expected by 2027 [12]. In October, the company broke ground on a molten salt production facility at its Manufacturing Development Campus in Albuquerque, New Mexico, with salt production scheduled for 2026 [13]. Finally, in November 2024, the NRC issued Construction Permits for Hermes 2, a two-unit test reactor facility featuring 35 MWt per unit, marking the first advanced nuclear plant designed to produce electricity [10]. The Hermes 2 reactors will be built using insights gained from the Hermes demonstration reactor.

\paragraph{Reference}

\begin{enumerate}
    \item \textbf{Kairos Power}. KP-FHR Core Design and Analysis Methodology Topical Report. 2024.
    \item \textbf{Kairos Power}. Technology Specifications. 2025.
    \item \textbf{Kairos Power}. Reactor Coolant for the Kairos Power Fluoride Salt-Cooled High Temperature Reactor Topical Report. 2020.
    \item \textbf{Kairos Power}. Safety Analysis Methodology for the Kairos Power Fluoride Salt-Cooled High-Temperature Test Reactor Topical Report. 2024.
    \item \textbf{Kairos Power}. Hermes 2 Non-Power Reactor Preliminary Safety Analysis Report. 2024.
    \item \textbf{Kairos Power}. KP-FHR Fuel Performance Methodology Topical Report. 2021.
    \item \textbf{U.S. Department of Energy}. Energy Department’s Advanced Reactor Demonstration Program Awards \$30 Million in Initial Funding for Risk Reduction Projects. 2020.
    \item \textbf{U.S. Nuclear Regulatory Commission}. Hermes – Kairos Application.
    \item \textbf{U.S. Department of Energy}. NRC Approves Construction for Hermes Reactor. 2023.
    \item \textbf{U.S. Nuclear Regulatory Commission}. Hermes 2 – Kairos Application.
    \item \textbf{Kairos Power}. U.S. Department of Energy and Kairos Power Execute Novel Performance-Based, Fixed-Price Milestone Contract. 2024.
    \item \textbf{U.S. Department of Energy}. Kairos Power Starts Construction of Hermes Reactor. 2024.
    \item \textbf{U.S. Department of Energy}. Kairos Power Breaks Ground on Molten Salt Production Facility. 2024.
\end{enumerate}

\clearpage
\newpage
\subsection{Natrium (TerraPower)}

\begin{wraptable}{o}{9cm} 
    \vspace{-12mm} 
    \centering
    \begin{threeparttable} 
        \renewcommand{\arraystretch}{1.2} 
        \rowcolors{2}{gray!15}{white} 
        \begin{tabular}{ p{4cm} | p{4cm} } 
            \toprule
            \rowcolor{white}
            \multicolumn{2}{l}{\textbf{General Information}} \\ 
            \midrule
            Reactor Type & Sodium-Cooled Fast Reactor \\
            Purpose & Commercial \\
            Thermal Power (MWt) & 840 \\
            Net Power Output (MWe) & 345 \\
            Design Life & 60 years \\
            Reactor Units per Site & - \\
            Seismic Design & 0.3 g \\ 
            Site Footprint (m\textsuperscript{2}) & - \\
            Construction Time (NOAK) & - \\
            
            \midrule
            \rowcolor{white}
            \multicolumn{2}{l}{\textbf{Fuel $\&$ Materials}} \\ 
            \midrule
            
            Core Coolant & Sodium \\
            Neutron Moderator & None \\
            Solid Burnable Absorber  & - \\
            Fuel Cladding & - \\
            Fuel Material & Metallic Uranium HALEU \\
            Fuel Enrichment & 19.75\% \\
            Refueling Cycle & 18 to 24 months \\
            
            \midrule
            \rowcolor{white}
            \multicolumn{2}{l}{\textbf{Development $\&$ Licensing}} \\ 
            \midrule
            
            Design Status & Detailed Design \\
            Licensing Status & -   
            \end{tabular}

        \begin{tablenotes}
            \item {Last ARIS update on 2024/11/21}
        \end{tablenotes}
    \end{threeparttable}
    \vspace{-5mm} 
\end{wraptable}

\paragraph{Basic Design} 
The Natrium advanced reactor is a pool-type, sodium-cooled fast reactor using high-assay low-enriched uranium zirconium fuel. It has a thermal capacity of 840 MWt and features a molten salt energy storage system, enabling flexible electricity generation and industrial applications [1,2].

\begin{figure}[h]
    \centering
    \includegraphics[width=0.8\linewidth]{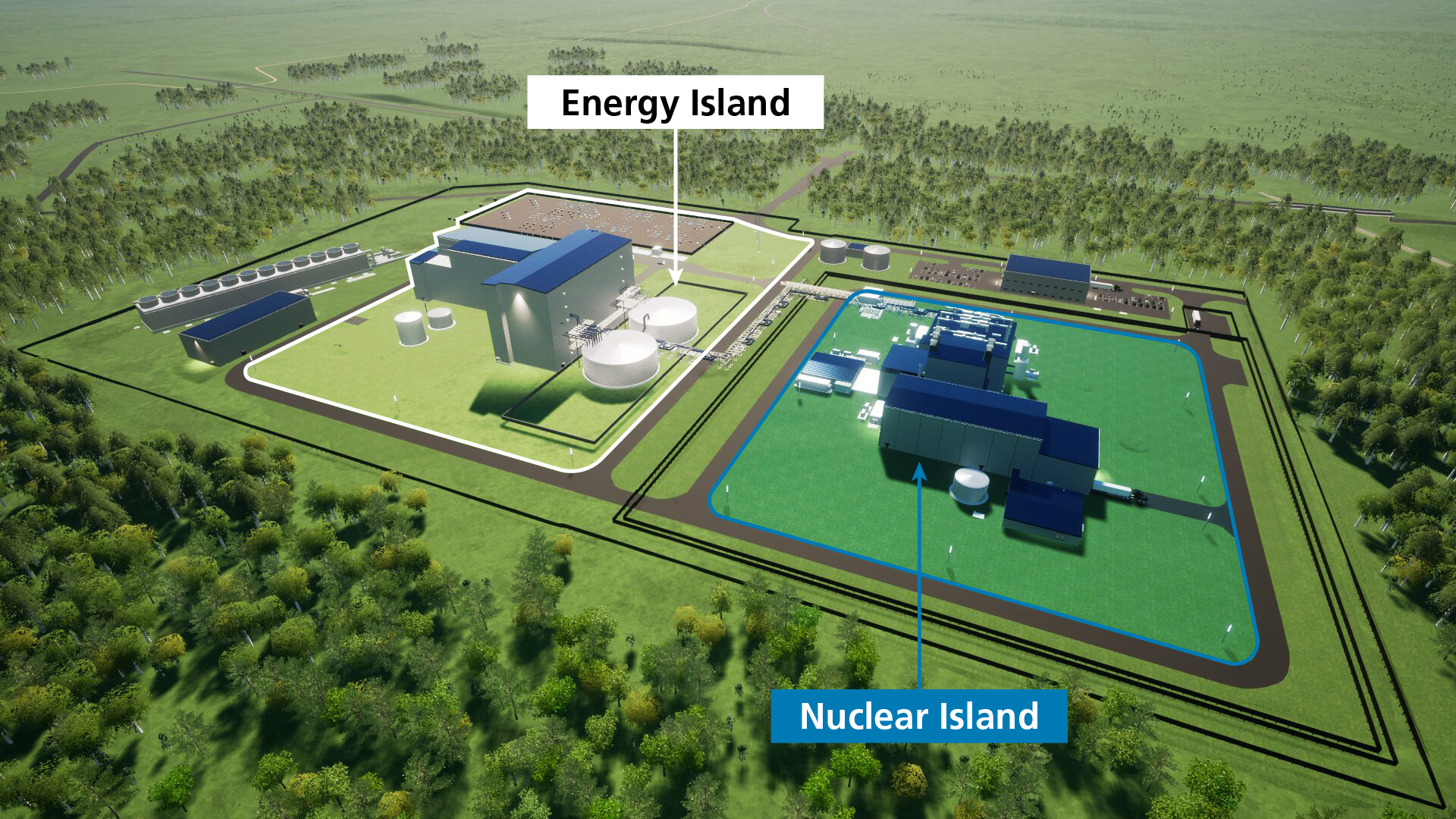}
    \caption{Rendering of the Natrium Plant design. Image Credit: TerraPower.}
    \label{fig:natrium-plant}
\end{figure}

\begin{figure}[h]
    \centering
    \includegraphics[width=0.7\linewidth]{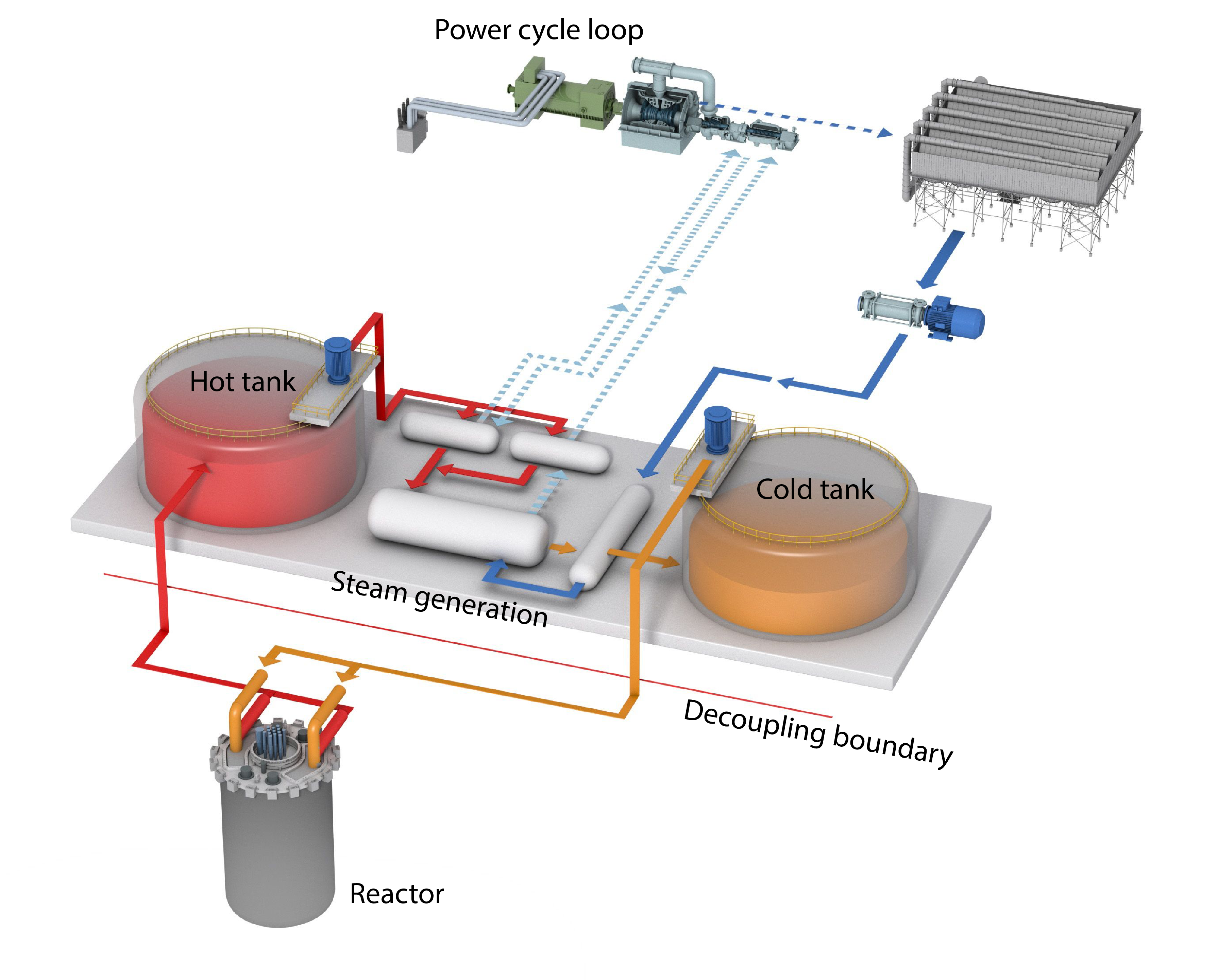}
    \caption{Components of the Nuclear Island and Energy Island of the Natrium Plant. Image Credit: TerraPower.}
    \label{fig:detailed-natrium-plant}
\end{figure}

A key design feature of Natrium is the decoupling of the Nuclear Island and Energy Island, as shown in Figures  \ref{fig:natrium-plant} and \ref{fig:detailed-natrium-plant}, ensuring the plant's transient separation and operational flexibility. The Nuclear Island, which houses the reactor and its support systems, operates independently of the Energy Island, where electricity is generated. This separation prevents Energy Island transients from affecting Nuclear Island safety. Additionally, molten salt storage tanks in the Energy Island decouple reactor operation from real-time turbine demand, allowing the Nuclear Island to operate continuously at full power with a capacity factor exceeding 90\%. The Energy Island can generate 338 MWe under steady-state conditions and ramp up to 500 MWe for limited durations to meet grid demands [2].

The reactor operates at near-atmospheric pressure, utilizing sodium’s high boiling point to eliminate the need for high-pressure containment. Heat is transferred from the primary sodium pool to the intermediate sodium loop through two Intermediate Heat Exchangers located within the reactor vessel. The heated sodium in the intermediate loop is then circulated by Intermediate Sodium Pumps to the sodium/salt heat exchangers, where it transfers energy to the molten salt coming from the cold tank in the Energy Island. The heated molten salt is subsequently stored in the hot tank for later power generation. Essentially, the Energy Island's design is similar to molten salt systems used in concentrated solar power plants, with the heat from the Natrium reactor replacing solar energy as the power source [1,2].

\paragraph{Fuel Design}
The Natrium plant will begin operation with Type 1 fuel, which closely resembles the fuel pins previously used in sodium fast reactors such as Experimental Breeder Reactor II and Fast Flux Test Facility. Type 1 fuel uses a metallic uranium-zirconium alloy (U-10 wt.\% Zr) as the fissile material, with uranium enrichment varying by core position and a peak enrichment of less than 20\% U-235 [3].

Each Type 1 fuel pin consists of a fuel cladding, upper and lower end caps, wire wrap, sodium-bonded fuel column, fission gas plenum, tag gas capsule, and axial shield. The stacked cylindrical fuel slugs are sodium-bonded and encapsulated in HT9 stainless steel cladding, with a nominal smear density of 75\%. Sodium bonding between the fuel and cladding ensures effective heat transfer, particularly early in life before the fuel expands to contact the cladding. Above the fuel column, a fission gas plenum, initially backfilled with inert gas, accommodates generated gaseous fission products during irradiation. The fuel pins are arranged in hexagonal fuel assemblies similar to those in Experimental Breeder Reactor II and Fast Flux Test Facility, with helical HT9 wire wraps maintaining pin-to-pin and pin-to-duct spacing while enhancing coolant mixing [3].

A unique aspect of the Natrium core includes the Lead Demonstration Assemblies and Lead Test Assemblies, which can be rapidly removed for post-irradiation examinations to support fuel performance evaluation and qualification efforts. Specifically, the Lead Demonstration Assemblies closely resemble standard Type 1 fuel assemblies and are used to generate data to reduce fuel performance uncertainties and extend burnup limits. The Lead Test Assembly program, while still under development, aim to qualify Type 1B fuel, which incorporates design advancements to achieve extended reactor residence times, higher burnup, and elevated coolant outlet temperatures, improving overall fuel cycle efficiency [3].

Fuel handling operations from the reactor core to the spent fuel pool are managed by three systems: the In-Vessel Fuel Handling System, the Ex-Vessel Fuel Handling System, and the Water Pool Fuel Handling System. The In-Vessel Fuel Handling System interfaces with the Ex-Vessel Fuel Handling System to transfer core assemblies out of the reactor vessel and into the Pool Immersion Cell, which facilitates the transition of core assemblies from a sodium-filled to a water-filled environment [4].

The Pool Immersion Cell is part of the Water Pool Fuel Handling System, which provides a water-based environment for intermediate-term storage of spent fuel and non-fuel irradiated core assemblies before disposal. Within the Pool Immersion Cell, core assemblies are placed in a cleaning vessel, where residual sodium coolant is chemically removed in a controlled setting to prevent sodium-water interactions. After cleaning and rinsing, the Pool Immersion Cell transports the core assembly underwater to the cask loading pit area of the spent fuel pool [4].

During normal operations, the spent fuel pool is actively cooled using two independent cooling trains that transfer heat to the Nuclear Island's Water System. If active cooling is lost, the spent fuel pool relies on evaporation for passive cooling, and redundant makeup water systems replenish its water inventory. When fully loaded, the spent fuel pool's water volume provides passive cooling for at least seven days without makeup water. When necessary, makeup water can be supplied via a connection to the Nuclear Island's Water System under normal operation conditions and from offsite sources through an external fill leg [4].

\paragraph{Reactivity Control}
Reactivity in the Natrium reactor is primarily controlled through control rods, which regulate neutron absorption to adjust power levels and facilitate reactor shutdown. The control rod drive mechanism moves the neutron absorber bundles within hexagonal control assemblies to control reactivity under normal operation and accident conditions. In response to a scram signal, the absorber bundle is released from the driveline and falls into the core by gravity, ensuring rapid shutdown [1,3]. 

To fulfill design diversity requirements, Natrium employs nine primary control rod assemblies and four secondary control rod assemblies [4,5]. The secondary control rod assembly is designed to provide geometric diversity to prevent common cause failures that could impair control rod operations. The primary and secondary control rod assemblies are expected to nearly identical, with key differences in the the number of absorber pins used, changes to the control assembly geometry, and the space between the inner duct and the guide tube [3]. 

The control rod absorber pins for both the primary and secondary control fuel assemblies consist of HT9 stainless steel cladding enclosing natural boron carbide pellets, which serve as the neutron-absorbing material [3,5]. The absorber pins have an upper plenum to manage the gaseous fission products released from the boron carbide and its axial expansion during irradiation. A pellet-to-clad gap is included to accommodate pellet swelling, which may reduce the fuel cycle length due to strain on the cladding [3].

\paragraph{Safety Features}
Natrium’s design is based on three fundamental safety functions: Retaining radionuclides, Controlling heat generation, and Controlling heat removal. Radionuclide retention is achieved through a functional containment strategy that relies on multiple passive barriers, starting from the radionuclide source and extending through all structures, systems, and components that separate it from the environment [4].

The fuel cladding serves as the first safety-related radionuclide retention barrier, providing mechanical support to withstand plenum gas pressure and fuel swelling due to fission product generation, as detailed in the Fuel Design section. Beyond the cladding, the Reactor Enclosure System functions as the next radionuclide retention barrier while also forming part of the primary coolant boundary [4].

The Reactor Enclosure System consists of the reactor vessel, reactor vessel head, reactor internal structures, Guard Vessel, and various support structures, all housed within the hardened reactor building to protect against external hazards. The reactor vessel contains an integrated primary heat transport system with no penetrations below the primary coolant operating level, reducing the risk of primary coolant loss. The reactor vessel is enclosed by the Guard Vessel, which acts as a redundant barrier for coolant containment and functional containment in the rare event of a reactor vessel leak. The annular gap between the reactor vessel and Guard Vessel is filled with argon gas to reduce the risk of sodium-air reactions [4]. 

Additional safety features of the Reactor Enclosure System include the low operating pressure of the primary coolant system, along with the sodium cover gas, which limits potential radionuclide transport. The intermediate sodium loop also operates at a higher pressure than the primary loop, ensuring that in the event of a leak between the primary and intermediate sodium boundary, sodium enters the primary system rather than allowing primary coolant to escape into the intermediate system [4].

The fundamental safety function of controlling heat generation is achieved by inserting control rods into the reactor core, either actively via control rod drive motors or passively through gravity insertion during a scram, as discussed in the Reactivity Control section. The Reactor Enclosure System also play a significant role in reactivity control by providing positional control and alignment to the control rod drive mechanisms [4].

The Intermediate Air Cooling system and Reactor Air Cooling system are responsible for removing reactor decay heat and dissipating it into the atmosphere. The Intermediate Air Cooling system provides both active and passive heat removal from the intermediate sodium loop, while the Reactor Air Cooling system operates continuously to remove heat from the Guard Vessel walls. Here, the Reactor Enclosure System plays a crucial role in maintaining the primary coolant flowpath, ensuring heat is transferred from the core to the intermediate sodium loop for the Intermediate Air Cooling system, and to the reactor vessel and Guard Vessel walls for the Reactor Air Cooling system. Notably, heat transfer occurs regardless of whether primary coolant flow is driven by forced circulation through the Primary Sodium Pumps or by natural circulation when these pumps are unavailable [4].

The Intermediate Air Cooling system transfers heat from the intermediate sodium loop to the atmosphere through a sodium-air heat exchanger, which is connected in series within the intermediate sodium loops. Under normal operating conditions, the Intermediate Air Cooling system serves as the primary and preferred method for reactor heat removal. It can function in three modes: active, blower, and passive. In active mode, both the Intermediate Sodium Pumps and Intermediate Air Cooling system's blower operate to provide forced circulation of intermediate sodium and airflow across the sodium-air heat exchanger. If the Intermediate Sodium Pumps are unavailable, blower mode ensures cooling by relying on natural circulation within the intermediate loop, while passive mode provides emergency cooling through natural circulation in the intermediate loop and natural convection across the sodium-air heat exchanger [4].

The elevation difference between the Intermediate Heat Exchangers in the reactor vessel, where heated sodium has lower density, and the sodium-air heat exchangers, where cooled sodium has higher density, generates the potential energy necessary for natural circulation flow. In addition, natural convection across the sodium-air heat exchanger is driven by the heated air rising and exiting through the air stack exhaust openings, while cool air enters through the low-elevation air intake. As a result, in passive mode, the Intermediate Air Cooling system operates without requiring automatic actuations or manual operator actions, and no electrical power is needed to initiate natural convection airflow across the sodium-air heat exchanger [4].

The Reactor Air Cooling system facilitates passive decay heat removal through natural convection airflow, continuously dissipating heat into the atmosphere without requiring automatic actuation or operator intervention. The process begins with heat stored in the primary sodium coolant being transported to the Guard Vessel, whose outer surface is directly exposed to the airflow path of the Reactor Air Cooling system. The Guard Vessel subsequently transfers decay heat to the Reactor Air Cooling system through a combination of thermal radiation and convection [4].

Natural convection airflow in the Reactor Air Cooling system is driven by buoyancy forces, as air in the riser annulus, positioned next to the Guard Vessel, heats up from reactor decay heat. As the temperature increases, the air becomes less dense, causing it to rise and escape into the atmosphere. The rising air simultaneously draws in cool ambient air through the inlet stack, directing it downward through the downcomer annulus, where it then flows into the riser annulus to be heated, sustaining the natural convection cycle [4].

The heat transfer rate in the Reactor Air Cooling system is influenced by the temperature of the Guard Vessel wall and increases significantly as its temperature rises, due to the direct relationship between thermal radiation heat transfer rate and surface temperature. As a result, the heat removal capacity of the Reactor Air Cooling system dynamically adjusts based on reactor conditions and the cooling demands of the primary coolant and reactor core. Under normal conditions, when reactor heat removal is primarily handled by the Intermediate Air Cooling system, the Guard Vessel's surface temperature remains low, reducing the heat removal rate of the Reactor Air Cooling system. However, if heat transfer via the Intermediate Air Cooling system becomes unavailable, the sodium coolant temperature rises, leading to a significant increase in thermal radiation heat transfer, first from the reactor vessel to the Guard Vessel, and then from the Guard Vessel to the Reactor Air Cooling system, enabling more heat to be dissipated into the atmosphere [4].

\paragraph{Development Timeline}
TerraPower’s Natrium reactor project has advanced through several significant milestones, primarily within the United States. In 2020, TerraPower was selected to participate in the DOE’s Advanced Reactor Demonstration Program, receiving $\$$80 million in initial funding to support the demonstration of its Natrium reactor and the development of a metal fuel fabrication facility [6].

In 2021, TerraPower revealed its plan to build the Natrium reactor near a retired coal plant in Kemmerer, Wyoming, and subsequently completed the land purchase on August 16, 2023 [7,8]. The company adopted a proactive approach by submitting a Construction Permit application to the NRC in March 2024 for its Kemmerer Power Station Unit 1, becoming the first developer to apply for a Construction Permit for a commercial advanced reactor [9]. The application was docketed by the NRC in May 2024, and by June 2024, non-nuclear construction activities were already underway at the Wyoming site [10,11].

In January 2025, TerraPower secured the first state permit for the Kemmerer Power Station Unit 1, allowing the company to proceed with construction and operational work not governed by the NRC [12]. Moving forward, TerraPower plans to focus on the construction of the Kemmerer Training Center and the energy island throughout 2025, with the NRC’s decision on the Construction Permit anticipated by the company in late 2026 [12].

\paragraph{Reference}
\begin{enumerate}
    \item \textbf{TerraPower}. Regulatory Management of Natrium Nuclear Island and Energy Island Design Interfaces, Topical Report. 2022.
    \item \textbf{TerraPower}. Energy Island Decoupling Strategy, White Paper. 2022.
    \item \textbf{TerraPower}. Fuel and Control Assembly Qualification, Topical Report. 2023.
    \item \textbf{TerraPower}. Kemmerer Power Station Unit 1 Preliminary Safety Analysis Report. 2024.
    \item \textbf{TerraPower}. Core Design and Thermal Hydraulic Technical Report. 2024.
    \item \textbf{U.S. Department of Energy}. U.S. Department of Energy Announces $\$$160 Million in First Awards under Advanced Reactor Demonstration Program. 2020.
    \item \textbf{U.S. Department of Energy}. Next-Gen Nuclear Plant and Jobs Are Coming to Wyoming. 2021.
    \item \textbf{TerraPower}. TerraPower Purchases Land in Kemmerer, Wyoming for Natrium Reactor Demonstration Project. 2023.
    \item \textbf{TerraPower}. TerraPower Submits Construction Permit Application to the U.S. Nuclear Regulatory Commission for the Natrium Reactor Demonstration Project. 2024.
    \item \textbf{U.S. Department of Energy}. NRC Dockets Construction Permit Application for TerraPower’s Natrium Reactor. 2024.
    \item \textbf{American Nuclear Society}. TerraPower breaks ground on SMR project in Wyoming. 2024.
    \item \textbf{TerraPower}. TerraPower Awarded Pivotal State Permit for Natrium Plant. 2025.
\end{enumerate}

\clearpage
\newpage
\subsection{eVinci (Westinghouse Electric Company)}

\begin{wraptable}{o}{8.5cm} 
    \vspace{-12mm} 
    \centering
    \begin{threeparttable} 
        \renewcommand{\arraystretch}{1.2} 
        \rowcolors{2}{gray!15}{white} 
        \begin{tabular}{ p{4cm} | p{3.5cm} } 
            \toprule
            \rowcolor{white}
            \multicolumn{2}{l}{\textbf{General Information}} \\ 
            \midrule
            Reactor Type & Other \\
            Purpose & Commercial \\
            Thermal Power (MWt) & 15 \\
            Net Power Output (MWe) & 5 \\
            Design Life & 8 years \\
            Reactor Units per Site & Single unit \\
            Seismic Design & 1 g \\ 
            Site Footprint (m\textsuperscript{2}) & $\sim$ 13,000 \\
            Construction Time (NOAK) & Less than 12 months \\
            
            \midrule
            \rowcolor{white}
            \multicolumn{2}{l}{\textbf{Fuel $\&$ Materials}} \\ 
            \midrule
            
            Core Coolant & Heat pipes (Na) \\
            Neutron Moderator & Graphite \\
            Solid Burnable Absorber  & - \\
            Fuel Cladding & TRISO \\
            Fuel Material & UCO TRISO \\
            Fuel Enrichment & 19.75\% \\
            Refueling Cycle & over 96 months\\
            
            \midrule
            \rowcolor{white}
            \multicolumn{2}{l}{\textbf{Development $\&$ Licensing}} \\ 
            \midrule
            
            Design Status & Detailed Design\\
            Licensing Status & -       
            \end{tabular}

        \begin{tablenotes}
            \item {Last ARIS update on 2024/10/17}
        \end{tablenotes}
    \end{threeparttable}
    \vspace{-8mm} 
\end{wraptable}

\paragraph{Basic Design}
The eVinci microreactor is a high-temperature heat pipe reactor capable of generating 15 MWt (5 MWe) and is designed to operate for over eight years at full power before refueling. Built for transportability and rapid deployment, eVinci requires minimal on-site installation and is compact enough to be transported via truck, rail, or waterway. It is well-suited for remote and off-grid applications, including powering remote communities, data centers, military installations, and cogeneration [1].

The reactor core consists of a graphite block with integrated fuel and heat pipe channels, powered by TRISO fuel. Unlike traditional reactors that rely on large volumes of coolant and active circulation, eVinci uses a passive heat transfer system that eliminates the need for pumps to force coolant circulation. Small amounts of sodium, contained within sealed heat pipes, transport heat from the core to the power conversion system, where it is converted into electricity using an open-air Brayton cycle [1].

\paragraph{Fuel Design}
eVinci's core consists of graphite blocks arranged in segmented, hexagonal unit cells that extend horizontally along the length of the core. These unit cells contain channels for fuel, burnable absorbers, heat pipes, and shutdown rods. The entire core assembly is housed within a sealed canister filled with inert gas at slightly above atmospheric pressure. This environment protects reactor components from oxidation while enhancing heat transfer. Surrounding the core is a thick radial reflector, which contains the control drums and plays a critical role in maintaining reactivity [2]. 

The microreactor uses High-Assay, Low-Enriched Uranium TRISO-coated fuel particles embedded in a graphite matrix, whose design has been discussed in the Xe-100 and KP-FHR sections [3]. Each TRISO-coated fuel particle contains a fissile fuel kernel surrounded by four protective layers: a porous carbon buffer layer, an inner pyrolytic carbon layer, a silicon carbide layer, and an outer pyrolytic carbon layer. These layers form multiple barriers that effectively retain fission products, significantly enhancing fuel safety. 

Unlike the spherical fuel elements used in the Xe-100 and KP-FHR reactors, eVinci employs TRISO fuel in the form of compact cylindrical pellets. Each fuel compact is created by pressing TRISO-coated particles into a right circular cylinder shape using a mixture of graphite powder and a binding agent. The graphite matrix enhances fuel performance by offering high-temperature strength, stability, and thermal conductivity, while also serving as a neutron moderator within the core [3].

eVinci is designed for an 8-year full-power operation, after which the entire spent core is transported in its original canister to a licensed facility. At this facility, the spent fuel is removed, stored in casks, and replaced with a newly fueled reactor core. This off-site refueling approach eliminates the need for on-site spent fuel storage and ensures safe handling within designated facilities until a permanent disposal location is available [4].

\paragraph{Reactivity Control}
Reactivity in the eVinci is monitored by power range and source range neutron detectors and is controlled by the Control Drum System and the Shutdown Rod System [1,2].

The Control Drum System consists of rotating control drums housed within the radial neutron reflector, which surrounds the core block. The control drums manage power levels by allowing absorber material to passively turn inward toward the core when power demand is reduced or lost, and turning a reflector material toward the core when demand increases [1,5]. The control drums provide reactivity control during normal operation and can shut down the reactor in case of power loss. The Shutdown Rod System provides an alternate means of shutdown through a passively actuated mechanism. This system offers an independent backup to the control drums for ensuring reactor shutdown when necessary [1,2].

\paragraph{Safety Features}
The eVinci microreactor is designed for resiliency and passive safety, with the entire reactor packaged within a secure canister and installed in a reinforced structure on-site. The reactor requires minimal on-site staff, with most operations, maintenance, and security managed via a remote monitoring station that allows off-site personnel to oversee reactor performance [4].

The Canister Containment Subsystem and TRISO-coated fuel form a multi-layered containment system that ensures safe operation and prevents radiological release. The Canister Containment Subsystem encases the entire core, providing supplemental containment, while additional structures inside and outside the Canister Containment Subsystem offer further retention. The TRISO-coated fuel itself serves as an intrinsic containment barrier, with multiple protective layers preventing the release of fission products even under extreme conditions [1,4]. 

Heat removal is accomplished using the Passive Heat Removal System which passively dissipates excess heat through buoyancy-driven air flow. Ambient air is channeled through the Canister Containment Subsystem, allowing heat to escape naturally to the environment. The system is designed to remove heat at a rate exceeding the core’s decay heat generation, ensuring the reactor remains in a stable, safe condition after shutdown [1].

\paragraph{Development Timeline}
In 2020, the eVinci microreactor was selected to participate in the DOE’s Advanced Reactor Demonstration Program, receiving $\$$9.3 million in total funding [6]. Since then, the company has maintained ongoing collaboration with the NRC through the submission of topical reports, technical white papers, and annual updates to its Pre-Application Regulatory Engagement Plan. In February 2023, Westinghouse filed a Notice of Intent to submit licensing documentation to both the NRC and the Canadian Nuclear Safety Commission for a joint technical review. The company is also actively participating in the Vendor Design Review process with the Canadian Nuclear Safety Commission [7,8].

In October 2023, Westinghouse was awarded a Front-End Engineering and Experiment Design contract by the DOE to support the development of a test reactor at Idaho National Laboratory [9]. The one-fifth-scale prototype will be tested at the DOME test bed operated by Idaho National Laboratory’s National Reactor Innovation Center. Westinghouse's eVinci completed the Front-End Engineering and Experiment Design phase in September 2024, with reactor testing projected to begin as early as 2026 [10].

More recently, in December 2024, Westinghouse achieved another key milestone when the NRC approved the eVinci's Instrumentation and Control platform, making eVinci the first microreactor design to receive NRC approval for its Instrumentation and Control system [11]. Early this year in January, Westinghouse announced an extension of its collaboration with NASA and the DOE to continue work on a space microreactor design under the Fission Surface Power project to develop compact, reliable power systems for space missions [12].

\paragraph{Reference}
\begin{enumerate}
    \item \textbf{Southern Company}. Westinghouse eVinci Micro-Reactor Tabletop Exercise Report. 2021.
    \item \textbf{Westinghouse Electric Company}. Nuclear Design Methodology Topical Report. 2024.
    \item \textbf{Westinghouse Electric Company}. Westinghouse TRISO Fuel Design Methodology Topical Report. 2024.
    \item \textbf{International Atomic Energy Agency}. Small Modular Reactor Technology Catalogue 2024 Edition, Second edition. 2025. 
    \item \textbf{Power Engineering}. Westinghouse sees a tech disrupter in its eVinci microreactor. 2022. 
    \item \textbf{U.S. Department of Energy}. Energy Department’s Advanced Reactor Demonstration Program Awards \$30 Million in Initial Funding for Risk Reduction Projects. 2020.
    \item \textbf{Westinghouse Electric Company}. Westinghouse Begins Joint Licensing Process with U.S. and Canadian Regulators for eVinci Microreactor. 2023.
    \item \textbf{Westinghouse Electric Company}. Westinghouse Begins Vendor Design Review for eVinci Microreactor with Canadian Nuclear Safety Commission. 2023.
    \item \textbf{Westinghouse Electric Company}. eVinci Microreactor Selected for Department of Energy FEEED Contract. 2023.
    \item \textbf{U.S. Department of Energy}. Westinghouse Completes Study for First eVinci Microreactor Experiment. 2024.
    \item \textbf{Westinghouse Electric Company}. Westinghouse eVinci Control System Achieves Major U.S. Licensing Milestone. 2024.
    \item \textbf{Westinghouse Electric Company}. Westinghouse Awarded NASA-DOE Contract to Continue Development of Space Microreactor Concept. 2025.
\end{enumerate}

\clearpage
\newpage
\section{Licensing and Regulatory Framework}
\subsection{NRC Licensing Process}
\label{section : nrc licensing}
The licensing regulations for domestic nuclear plants are outlined in Title 10 of the Code of Federal Regulations. Currently, the NRC provides two licensing pathways for nuclear reactors: Part 50, which follows a two-step process with separate construction and operating licenses, and Part 52, which uses a one-step process that combines both approvals. A third pathway, Part 53, is under development to streamline licensing for advanced reactors, though it is not yet finalized.

Prior to submitting a formal application under Part 50 or Part 52, companies are expected to engage in pre-application activities with the NRC. This typically begins with a Letter of Intent, outlining the project’s scope and timeline [1]. A Regulatory Engagement Plan is also encouraged to communicate planned pre-application activities, as seen in submissions by Westinghouse and Holtec for the AP300 and SMR-300, respectively [2]. 

Pre-application activities include NRC meetings and the submission of topical reports and white papers, detailing key technical aspects of the design [3]. The applicant is also expected to provide a draft application for pre-application readiness assessment, allowing both parties to address potential gaps before the formal application submission [3,4]. Additionally, the NRC holds a public hearing near the proposed deployment site to engage stakeholders and address concerns.

The following sections outline the key components of the Part 50 and Part 52 licensing pathways, along with a brief discussion on the ongoing development of Part 53.

\subsubsection*{Part 50 Pathway}
The 10 CFR Part 50 pathway is a two-step licensing process that was originally introduced in 1956, which requires obtaining a Construction Permit, followed by an Operating License before operation.

To obtain a Construction Permit, the applicant must submit a Preliminary Safety Analysis Report and an Environmental Report, which undergo safety and environmental reviews by the NRC [5,6]. These documents evaluate the safety and health risks of the proposed design to workers and the general public, as well as the environmental impact of the facility. Only after NRC approval and mandatory hearings will the Construction Permit be issued, allowing the applicant to begin construction [7].

\begin{figure}[h]
    \centering
    \includegraphics[width=\linewidth]{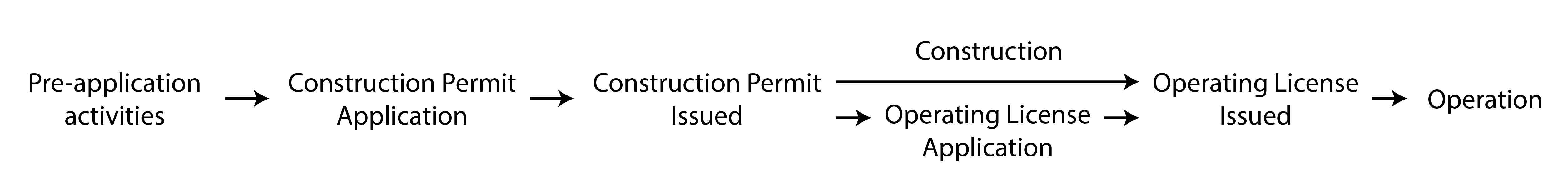}
    \caption{Simplified two-step licensing process under 10 CFR Part 50.}
    \label{fig:part_50}
\end{figure}

Once construction begins, the applicant must finalize its design and apply for an Operating License. This requires submitting a Final Safety Analysis Report and any necessary updates to the Environmental Report [5,8], along with other required documentation such as proposed Technical Specifications (\S 50.36), Emergency Plans (\S 50.47), and a Fire Protection Plan (\S 50.48). Throughout the construction phase, the NRC continues reviewing the Operating License application. Once construction is complete and all NRC regulatory concerns have been resolved, the Operating License can be issued, allowing the plant to begin operations. A simplified schematic adapted from the NRC is provided in Figure \ref{fig:part_50} for clarity.

The Part 50 licensing pathway provides flexibility in modifying the reactor design, as it does not need to be finalized until the Operating License application. Recent SMR projects such as those by TerraPower, Kairos Power, and X-energy\footnote{According to the NRC, X-energy is planning to apply for a Construction Permit.} have opted for this licensing approach, allowing companies to make significant progress while refining their final design before the Operating License application. More specifically, TerraPower submitted its Construction Permit application in March 2024 [9], with Construction Permits for Hermes 1 and 2 (Kairos Power) already issued by the NRC in 2023 and 2024, respectively [10,11]. 

However, this flexibility comes with risks, as the lack of a fully finalized design during construction may lead to challenges in the Operating License application stage. The uncertain nature of this design-as-you-go approach has also been criticized to result in escalating costs and delays in plant operations, leading to the proposal of a one-step licensing pathway in Part 52 [12].

\subsubsection*{Part 52 Pathway}
\begin{figure}[h]
    \centering
    \includegraphics[width=0.9\linewidth]{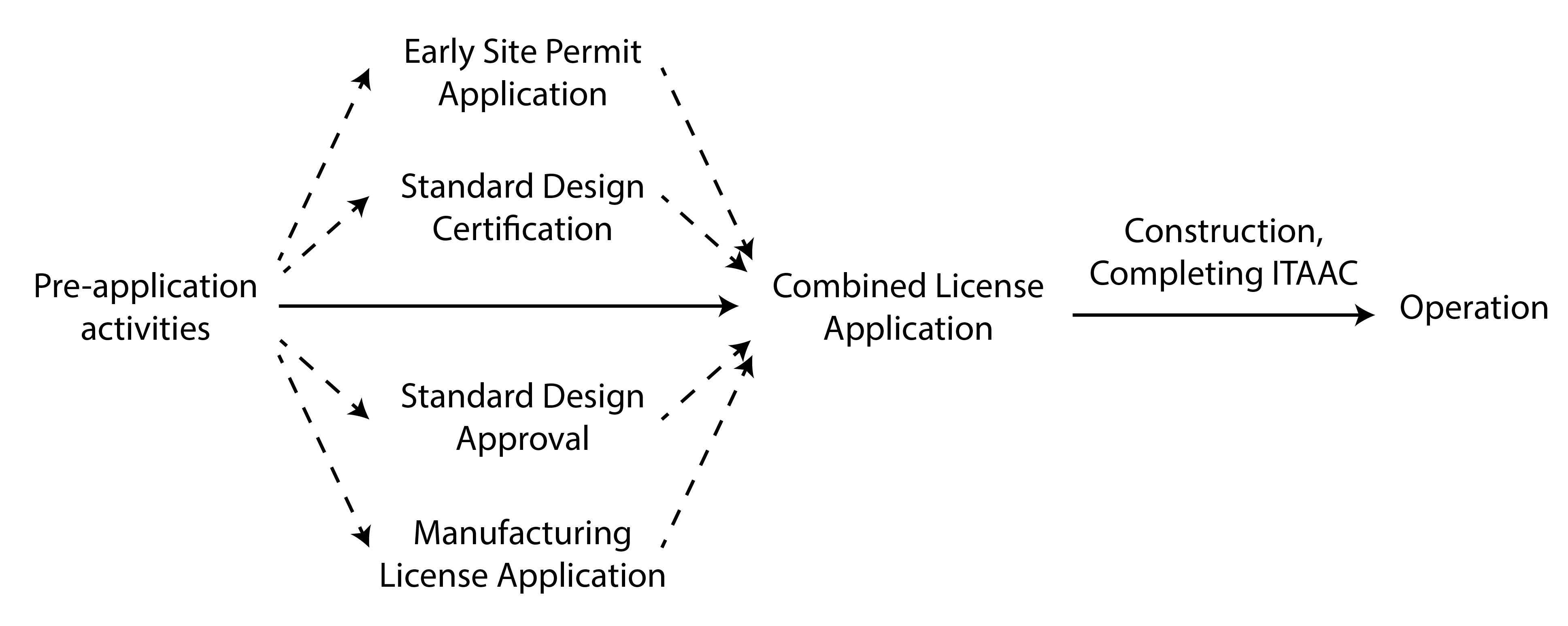}
    \caption{Simplified one-step licensing process under 10 CFR Part 52.}
    \label{fig:part_52}
\end{figure}

The Part 52 licensing pathway, introduced in 1989, provides a standardized and streamlined one-step process for nuclear reactor licensing. A Combined License authorizes both construction and conditional operation of a plant, eliminating the need for a separate Operating License. While not required, a Combined License application may reference previously issued approvals, such as an Early Site Permit, Standard Design Certification, Standard Design Approval, or Manufacturing License, which can simplify the licensing process [13]. A simplified licensing roadmap under Part 52 is shown in Figure \ref{fig:part_52}.

The Early Site Permit approves a nuclear facility site and remains valid for 10 to 20 years but does not authorize construction or operation [14] . In contrast, a Manufacturing License allows a nuclear reactor to be manufactured at a separate location before being installed at a site licensed under a Construction Permit (Part 50) or a Combined License (Part 52). A Manufacturing License application may reference a Standard Design Certification or Standard Design Approval for a specific plant design [15].

For reactor design approvals, a Standard Design Certification applies to the entire nuclear power plant design, while a Standard Design Approval can cover either the full design or a major portion of it [16,17]. The key difference is that Standard Design Certification is established through a rulemaking process and codified as an appendix to Part 52, providing greater regulatory stability [18]. In contrast, a Standard Design Approval offers less regulatory certainty, as the NRC can reassess it if new safety concerns arise [19]. The Standard Design Certification application process is more rigorous, requiring a Final Safety Analysis Report, Environmental Report, and proposed Inspections, Tests, Analyses, and Acceptance Criteria, whereas an Standard Design Approval application generally only requires an Final Safety Analysis Report [16,17]. 

Finally, a Combined License is required before construction can begin. The plant is also only allowed to operate once all identified Inspections, Tests, Analyses, and Acceptance Criteria are completed and verified by the NRC, ensuring that the plant has been constructed correctly and will operate as intended [20]. Since all required information must be provided and reviewed by the NRC before construction can begin under Part 52, a Combined License provides more finality compared to a Construction Permit issued under Part 50. As of March 2025, eight new reactor units with GEN III+ designs hold a Combined License, with Vogtle Units 3 and 4 being the only operational units, both using the AP1000 design [21]. 

Among the SMR designs discussed in Chapter 2, NuScale Power’s original US600 (12-module, 50 MWe) design remains the only SMR with an approved Standard Design Certification, while its updated UDS460 (6-module, 77 MWe) design is currently undergoing Standard Design Approval review [22,23]. As of March 2025, Oklo Power is the only company to have submitted a Combined License application for a non-LWR design, the Aurora fast microreactor, but its application was denied in 2022 [24].

\subsubsection*{Part 53 Pathway}
The Nuclear Energy Innovation and Modernization Act (NEIMA), enacted on January 14, 2019, mandated the NRC to establish a technology-inclusive, risk-informed, and performance-based regulatory framework for licensing advanced nuclear reactors. This requirement resulted in the development of 10 CFR Part 53, which serves as an alternative licensing pathway alongside 10 CFR Parts 50 and 52. NEIMA set a deadline for completing this rulemaking before December 31, 2027 [25]. 

The proposed Part 53 retains the same types of licenses available under Parts 50 and 52, such as construction permits, operating licenses, early site permits, design certifications, and combined licenses. However, unlike Parts 50 and 52, which were primarily developed around light-water reactor technology, Part 53 is designed to be technology-inclusive, accommodating a wider range of reactor designs, including non-LWRs [26]. 

On March 1, 2023, the NRC staff submitted a proposed Part 53 rule (SECY–23–0021), containing two frameworks: Framework A and Framework B, to the Commission for review. Framework A focuses on utilizing Probabilistic Risk Assessment to evaluate the likelihood and consequences of various accident scenarios, while Framework B was originally introduced in response to stakeholder feedback advocating for greater alignment with international standards. However, Framework B evolved into a traditional deterministic and prescriptive approach and was ultimately rejected by the Commission for inclusion in Part 53 in March 2024 [27,28,29].

The latest version of the proposed rule was published in the Federal Register on October 31, 2024, and the period for public comment closed on February 28, 2025.

\subsection*{Reference}
\begin{enumerate}
    \item \textbf{U.S. Nuclear Regulatory Commission}. Pre-application Process. 2025.
    \item \textbf{U.S. Nuclear Regulatory Commission}. Regulatory Engagement Plan. 2025.
    \item \textbf{U.S. Nuclear Regulatory Commission}. Review of Risk-Informed, Technology-Inclusive Advanced Reactor Applications—Roadmap, Appendix A. 2024.
    \item \textbf{Office of Nuclear Reactor Regulation}. Preapplication Readiness Assessment. 2020.
    \item \textbf{U.S. Nuclear Regulatory Commission}. 10 CFR \S 50.34 Contents of applications; technical information. 
    \item \textbf{U.S. Nuclear Regulatory Commission}. 10 CFR \S 51.50 Environmental report—construction permit, early site permit, or combined license stage. 
    \item \textbf{U.S. Nuclear Regulatory Commission}. 10 CFR \S 50.58 Hearings and report of the Advisory Committee on Reactor Safeguards.
    \item \textbf{U.S. Nuclear Regulatory Commission}. 10 CFR \S 51.53 Postconstruction environmental reports.
    \item \textbf{U.S. Nuclear Regulatory Commission}. TerraPower, LLC -- Kemmerer Power Station Unit 1 Application. 2025.
    \item \textbf{U.S. Nuclear Regulatory Commission}. Hermes – Kairos Application. 2024.
    \item \textbf{U.S. Nuclear Regulatory Commission}. Hermes 2 – Kairos Application. 2024.    
    \item \textbf{Stephen G. Burns}. Looking Backward, Moving Forward: Licensing New Reactors in the United States. 2008.
    \item \textbf{U.S. Nuclear Regulatory Commission}. 10 CFR \S 52.73 Relationship to other subparts.
    \item \textbf{U.S. Nuclear Regulatory Commission}. 10 CFR \S 52.26 Duration of permit.
    \item \textbf{U.S. Nuclear Regulatory Commission}. 10 CFR \S 52.153 Relationship to other subparts.
    \item \textbf{U.S. Nuclear Regulatory Commission}. 10 CFR \S 52.47 Contents of applications; technical information.
    \item \textbf{U.S. Nuclear Regulatory Commission}. 10 CFR \S 52.137 Contents of applications; technical information.
    \item \textbf{U.S. Nuclear Regulatory Commission}. 10 CFR \S 52.63 Finality of standard design certifications.
    \item \textbf{U.S. Nuclear Regulatory Commission}. 10 CFR \S 52.145 Finality of standard design approvals; information requests.
    \item \textbf{U.S. Nuclear Regulatory Commission}. Inspections, Tests, Analyses, and Acceptance Criteria (ITAAC). 2022.
    \item \textbf{U.S. Nuclear Regulatory Commission}. Combined License Holders for New Reactors. 2025.
    \item \textbf{U.S. Department of Energy}. NRC Certifies First U.S. Small Modular Reactor Design. 2023.
    \item \textbf{U.S. Nuclear Regulatory Commission}. NuScale US460 Standard Design Approval Application Review. 2025.
    \item \textbf{U.S. Nuclear Regulatory Commission}. Aurora – Oklo Application. 2022.
    \item \textbf{U.S. Congress}. Nuclear Energy Innovation and Modernization Act. 2019.
    \item \textbf{U.S. Nuclear Regulatory Commission}. Risk-Informed, Technology-Inclusive Regulatory Framework for Advanced Reactors. 2024.
    \item \textbf{U.S. Nuclear Regulatory Commission}. SRM-SECY-23-0021. 2024.
    \item \textbf{Power Magazine}. NRC Sets Stage for Advanced Nuclear with New Part 53 Rule. 2024.
    \item \textbf{The Breakthrough Institute}. Nuclear Regulatory Commission Charts a Path Forward on Part 53. 2024.    
\end{enumerate}

\clearpage
\newpage
\subsection{Notable Policies and Programs}
Various legislative support and government funded programs are indispensable to the rapid development of SMR technology in the U.S. in recent years. Here, a brief summary of several recent policies and programs, their scope, contents, and targets, are provided.

\paragraph{2015 - Gateway for Accelerated Innovation in Nuclear} 
The Gateway for Accelerated Innovation in Nuclear (GAIN) is a DOE initiative launched in 2015 to connect industry with national laboratories and accelerate the development of advanced nuclear technologies. It serves as a central access point to the expertise, facilities, and research capabilities within the national lab system.

A key component of GAIN is the Nuclear Energy Voucher Program, which offers up to \$500,000 per project to  to support industry collaboration with national labs. Since its inception, GAIN has awarded over 100 vouchers, totaling nearly \$40 million [1]. In addition to funding, GAIN facilitates collaborative agreements, enabling private entities to engage in research and development with national labs. It also offers regulatory guidance, helping companies navigate licensing and deployment challenges in the nuclear sector [2].

\paragraph{2016 - Licensing Modernization Project}
The Licensing Modernization Project (LMP) was an industry-led, DOE-supported initiative (2016–2019) aimed at modernizing the U.S. regulatory framework to better accommodate advanced non-light-water reactor concepts. Conventional NRC licensing processes were originally designed for large light-water reactors and made it difficult for novel reactor concepts to fit within existing regulatory requirements [3].

The LMP initiative resulted in the guidance document "Risk-Informed Performance-Based Technology Inclusive Guidance for Non-Light Water Reactor Licensing Basis Development" (NEI 18-04), which provides detailed guidance on the selection of license basis events, classification of structures, systems, and components, as well as assessment of defense-in-depth capabilities to demonstrate the safety and integrity of their reactor design [3,4]. In 2020, the NRC formally endorsed NEI 18-04 through Regulatory Guide 1.233, making it an officially approved licensing approach for non-LWRs under 10 CFR Part 50 and Part 52 [5].

\paragraph{2019 - Nuclear Energy Innovation and Modernization Act}
The Nuclear Energy Innovation and Modernization Act (NEIMA) was signed into law in January 2019 to modernize the regulatory framework for nuclear energy, particularly for advanced nuclear reactors [6]. Section 103 of NEIMA focuses specifically on regulations and licensing processes, directing the NRC to improve and develop new approaches for licensing advanced reactors.

These efforts include:
\begin{itemize}
    \item Establishing stages in the licensing process to enhance predictability and efficiency.
    \item Expanding the use of risk-informed, performance-based licensing evaluations.
    \item Developing strategies for licensing research and test reactors within the existing regulatory framework.
\end{itemize}

Most notably, NEIMA mandates the NRC to complete a new rulemaking by December 31, 2027, to establish a technology-inclusive regulatory framework for licensing commercial advanced reactors. This directive led to the development of Part 53 under 10 CFR, as explained in Section \ref{section : nrc licensing}.

\paragraph{2020 - Advanced Reactor Demonstration Program}
Launched by the DOE, the Advanced Reactor Demonstration Program (ARDP) aims to accelerate the deployment of advanced nuclear reactors through cost-shared partnerships with U.S. industry [7]. The program is structured into three tracks, each addressing different stages of reactor development:

\begin{itemize}
    \item Advanced Reactor Demonstrations [8]
    \begin{itemize}
        \item Supports fully functional reactor deployments within seven years.
        \item Recipients: TerraPower (Natrium reactor) and X-energy (Xe-100 reactor).
        \item The DOE plans to invest \$3.2 billion over seven years, with industry cost-sharing.
    \end{itemize}
    \item Risk Reduction for Future Demonstrations [9]
    \begin{itemize}
        \item Focuses on designs tackling key technical, regulatory, and operational issues to prepare for future deployment.
        \item Recipients: Kairos Power (Hermes reactor), Westinghouse (eVinci microreactor), BWXT (BWXT Advanced Nuclear Reactor), Holtec (SMR-160), and Southern Company Services (Molten Chloride Reactor Experiment).
        \item DOE expects to invest \$600 million over seven years.
    \end{itemize}
    \item Advanced Reactor Concepts 2020 [10]
    \begin{itemize}
        \item Supports innovative and early-phase reactor designs with commercialization potential in the 2030s.
        \item Recipients: Advanced Reactor Concepts (Inherently Safe Advanced SMR for American Nuclear Leadership), General Atomics (Fast Modular Reactor Conceptual Design), and MIT (Horizontal Compact High Temperature Gas Reactor).
        \item DOE expects to invest \$56 million over four years.
    \end{itemize}
\end{itemize}

\paragraph{2021 - Infrastructure Investment and Jobs Act}
The Infrastructure Investment and Jobs Act (IIJA), signed into law in 2021, includes provisions under Subtitle C — Nuclear Energy Infrastructure to support nuclear energy as a key component of decarbonization. These measures focus on both advancing new reactor technologies and preserving existing nuclear plants.

Section 40321 directs the DOE to assess the role of micro-reactors and SMRs in enhancing energy resilience and reducing carbon emissions. This includes plans for their deployment in remote communities and DOE facilities. Additionally, the DOE is tasked with providing financial and technical assistance for feasibility studies to identify suitable locations for advanced reactor deployment [11].

Another major provision is the Civil Nuclear Credit Program under Section 40323, which allocates \$6 billion to prevent the premature closure of economically struggling nuclear power plants, ensuring their continued contribution to clean energy goals [11]. The first Civil Nuclear Credits, totaling \$1.1 billion, was granted to Diablo Canyon Power Plant in California on January 17, 2024 [12].

\paragraph{2022 - Inflation Reduction Act}
The Inflation Reduction Act (IRA), signed into law in August 2022, provides various tax incentives to support both the operation of existing nuclear power plants and the deployment of advanced reactors for electricity generation. 

Specifically, the Zero-Emission Nuclear Power Production Credit (Section 45U) provides a per-kilowatt-hour tax credit to support the continued operation of existing nuclear facilities. Meanwhile, the Clean Electricity Production Credit (Section 45Y) and the Clean Electricity Investment Credit (Section 48E) are technology-neutral incentives for zero-emission power plants placed in service from 2025, promoting the deployment and commercial operation of advanced reactor such as SMRs [13].

In addition to tax credits, the IRA provides \$150 million to support general nuclear projects under the Office of Nuclear Energy. It also allocates \$700 million for the development of High-Assay Low-Enriched Uranium (HALEU), a critical fuel for many advanced reactor designs, under Section 50173 [13].

\paragraph{2024 - Accelerating Deployment of Versatile, Advanced Nuclear for Clean Energy Act}
The Accelerating Deployment of Versatile, Advanced Nuclear for Clean Energy (ADVANCE) Act, signed into law in 2024, introduces comprehensive reforms to facilitate the deployment of nuclear energy. The act focuses on streamlining licensing processes, reducing regulatory burdens, and promoting advanced nuclear fuel development [14]. Notable provisions include:

\begin{itemize}
    \item Section 201: Reduces licensing fees for advanced nuclear reactor applicants.
    \item Section 202: Provides awards for the first advanced reactors to achieve key regulatory milestones, such as obtaining an Operating License or Combined License.
    \item Section 203: Requires the NRC to address challenges in licensing advanced reactors for non-electric applications.
    \item Section 206: Directs the NRC to streamline licensing for nuclear reactors at retired coal plant sites.
    \item Section 207: Expedites Combined License applications for reactors that meet specific qualifications.
    \item Section 208: Requires the NRC to develop risk-informed, performance-based licensing guidelines and apply them under existing frameworks or new rulemaking.
\end{itemize}

Other key provisions address issues such as environmental reviews for nuclear plants (Section 506) and the development and licensing of advanced nuclear fuels (Section 404). 

\paragraph{2024 - Generation III+ Small Modular Reactor Program}
The DOE launched the Generation III+ Small Modular Reactor Program in October 2024, offering up to \$900 million to accelerate the initial deployment of Gen III+ SMR technologies. The funding is divided into two tiers: Tier 1 (up to \$800 million) focuses on assisting in the deployment of the first Gen III+ SMR plant and establishing a multi-reactor orderbook to drive further adoption. Tier 2 (up to \$100 million) supports efforts to address key barriers such as licensing, supply chain development, and site preparation for future expansion. Applications closed on January 17, 2025 [15].

\subsection*{Reference}
\begin{enumerate}
    \item \textbf{Gateway for Accelerated Innovation in Nuclear}. GAIN NE VOUCHERS.
    \item \textbf{Gateway for Accelerated Innovation in Nuclear}. Partnership Support.
    \item \textbf{Idaho National Laboratory}. Licensing Modernization Project for Advanced Reactor Technologies: FY 2018 Project Status Report. 2018.
    \item \textbf{Nuclear Energy Institute}. NEI 18-04, Rev. 1 Risk-Informed Performance-Based Technology Inclusive Guidance for Non-Light Water Reactor Licensing Basis Development. 2019.
    \item \textbf{U.S. Nuclear Regulatory Commission}. Regulatory Guide 1.233. 2020.
    \item \textbf{U.S. Congress}. Nuclear Energy Innovation and Modernization Act. 2019.
    \item \textbf{U.S. Department of Energy}. Advanced Reactor Demonstration Program.
    \item \textbf{U.S. Department of Energy}. U.S. Department of Energy Announces \$160 Million in First Awards under Advanced Reactor Demonstration Program. 2020.
    \item \textbf{U.S. Department of Energy}. Energy Department’s Advanced Reactor Demonstration Program Awards \$30 Million in Initial Funding for Risk Reduction Projects. 2020.
    \item \textbf{U.S. Department of Energy}. Energy Department’s Advanced Reactor Demonstration Program Awards \$20 million for Advanced Reactor Concepts. 2020.
    \item \textbf{U.S. Congress}. Infrastructure Investment and Jobs Act. 2021.
    \item \textbf{U.S. Department of Energy}. Biden-Harris Administration Finalizes Award of \$1.1 Billion in Credits to Pacific Gas and Electric’s Diablo Canyon Power Plant. 2024.
    \item \textbf{U.S. Congress}. Inflation Reduction Act. 2022.
    \item \textbf{U.S. Congress}. Accelerating Deployment of Versatile, Advanced Nuclear for Clean Energy Act. 2024.
    \item \textbf{U.S. Department of Energy}. Funding Notice: Generation III+ Small Modular Reactor Program. 2024.
\end{enumerate}

\clearpage
\newpage
\section*{Disclaimer}
This technical review on the development of SMRs in the U.S. compiles publicly available information from sources including the Nuclear Energy Agency, the International Atomic Energy Agency, U.S. Nuclear Regulatory Committee, and official websites of nuclear power plant operators. While we strive for accuracy, we cannot guarantee the completeness or correctness of all reported values. Additionally, to maintain conciseness and focus, some legislative details and technical specifics have been simplified. For a more comprehensive and detailed understanding, readers are encouraged to refer to the original documents. Any discrepancies should be verified with the original sources. The authors and affiliated institutions assume no responsibility for errors, omissions, or interpretations based on this document.







\end{document}